\documentclass[12pt,a4paper]{article}
\usepackage[a4paper, total={170mm,257mm}, left=20mm, top=20mm,]{geometry}
\usepackage[utf8]{inputenc}
\usepackage{graphicx}
\usepackage{dcolumn}
\usepackage{bm}
\usepackage{amsmath,amsfonts,amssymb,setspace}
\usepackage{mathrsfs}
\usepackage{mathtools}
 
\usepackage{slashed}
\usepackage{braket,xcolor}
\usepackage{verbatim}
\usepackage{caption}
\usepackage{subcaption}
\usepackage{multirow}
\usepackage{amsfonts}
\usepackage[utf8]{inputenc}
\usepackage{setspace, hyperref}
\usepackage{cite}
\usepackage{xcolor}

\usepackage[utf8]{inputenc}
\usepackage{braket}

\usepackage{hyperref}
\usepackage{mathrsfs}
\usepackage{graphicx}
\usepackage{bm}
\usepackage{braket}
\usepackage{hyperref}
\usepackage{color}
\usepackage{makecell}
\usepackage{enumitem}
\usepackage{upgreek}
\usepackage{makecell}

\usepackage{hyperref}
\usepackage{braket}
\usepackage{bm}
\usepackage{longtable}
\usepackage[bb=boondox]{mathalfa}
\usepackage{adjustbox}
\usepackage{array}
\usepackage{multirow}
\usepackage{tabularx}

\usepackage{float}

\usepackage{tikz}

\numberwithin{equation}{section}

\hypersetup{colorlinks=false, linkcolor=blue, citecolor=red}
\begin{document}

	\newcommand{\eqn}[1]{(\ref{#1})}

	\newcommand{\vs}[1]{\vspace{#1 mm}}
	\newcommand{\dsl}{\pa \kern-0.5em /}

	\onehalfspacing
	\parskip 0.03in
	\begin{flushright}
		%
	\end{flushright}
	\begin{center}
		{\Large{\bf   Q-curvature and Path Integral Complexity }}
		
		\vs{18}
		
		{Hugo A. Camargo${}^{a,b}$\footnote{hugo.camargo@aei.mpg.de}, Pawel Caputa${}^{c}$\footnote{pawel.caputa@fuw.edu.pl}, Pratik Nandy${}^{d}$\footnote{pratiknandy@iisc.ac.in}} 
		
		\vs{10}
		
		{\it ${}^{a}$Max-Planck-Institut für Gravitationsphysik, \\
			Am Mühlenberg 1, 14476 Potsdam-Golm, Germany}
		\vskip .5mm
		{\it ${}^{b}$Dahlem Center for Complex Quantum Systems,\\ Freie Universität Berlin, Arnimallee 14, 14195 Berlin, Germany}
		\vskip .5mm
		{\it ${}^{c}$Faculty of Physics, University of Warsaw, ul. Pasteura 5, 02-093 Warsaw, Poland}\vskip .5mm
		{\it ${}^{d}$Centre for High Energy Physics, Indian Institute of Science,\\ C.V. Raman Avenue, Bangalore-560012, India.}\vskip .5mm

	\end{center}
	
	\vs{5}
	
	\begin{abstract}
	
		We discuss the interpretation of path integral optimization as a uniformization problem in even dimensions. This perspective allows for a systematical construction of the higher-dimensional path integral complexity in holographic conformal field theories in terms of Q-curvature actions.  We explore the properties and consequences of these actions from the perspective of the optimization programme, tensor networks and penalty factors. Moreover, in the context of recently proposed holographic path integral optimization, we consider higher curvature contributions on the Hartle-Hawking bulk slice and study their impact on the optimization as well as their relation to Q-curvature actions and finite cut-off holography.
		
	\end{abstract}
	\newpage
	\tableofcontents

	\section{Introduction and Summary}
	\label{sec:intro}
In the past years, insights from quantum information have led to an abundance of results in the context of the AdS/CFT correspondence, also known as holography~\cite{Maldacena:1997re,Witten:1998qj}. While undoubtedly the main quantity in this story has been the notion of entanglement and its entropy~\cite{Ryu:2006bv}, the concept of complexity has become increasingly prominent in the field (see e.g.~review~\cite{Chen:2021lnq}). The importance of complexity in holography was motivated by the observation~\cite{Susskind:2014moa,Susskind:2014rva,Stanford:2014jda,Brown:2015bva} that co-dimension-one boundary-anchored maximal volumes and co-dimension-zero boundary-anchored causal developments, which appear to be natural probes of the black hole interior in holography, share similar properties with linearly growing tensor networks describing states dual to these black-hole spacetimes \cite{Hartman:2013qma}.
	
 As an information-theoretic quantity, the complexity of an operator or a state can be intuitively defined as the minimum number of quantum gates required to build the operator or to produce the state (see e.g.~\cite{Nielsen1133}). However, unlike in the case of entanglement entropy, a precise notion of complexity that would be universally applicable to quantum field theories (QFTs) and useful for holography is not obvious and still under very active development (see e.g. review~\cite{Chapman:2021jbh}). 
 In this work we will focus on a particular approach to state complexity, known as path integral complexity~\cite{Caputa:2017urj,Caputa:2017yrh}, that was developed purely in the language of Euclidean path integrals in QFT.  Its main idea is inspired by tensor networks (TN)~\cite{Miyaji:2016mxg,Vidal:2007hda} and tensor network renormalisation (TNR)~\cite{TNR,TNRmera} and regards the geometry on which the Euclidean path integral prepares a state in QFT as a specific TN. 
 
 More precisely, starting from the original metric (continuous TN) on which the path integral is computed (which is usually taken as Euclidean flat) and keeping the boundary conditions fixed, one deforms the metric to an arbitrary curved one. For example in conformal field theories (CFTs), our main focus in this work, the state prepared by the path integral over the new geometry is proportional to the one prepared in the flat geometry. The proportionality factor captures the amount of ``unnecessary computation", and its minimization leads to the optimal geometry (optimal TN), which turns out to be hyperbolic. Physically, this means that instead of performing the path integral over a flat geometry, the ``cost” of preparing the state will be minimal if one performs it over hyperbolic space/TN. In $2$-dimensions, this cost functional has been identified with the Liouville action and the on-shell value of this action yields the measure of path integral complexity. More operational aspects of the Liouville action and its generalizations were further discussed in~\cite{Czech:2017ryf,Caputa:2018kdj,Camargo:2019isp,Bhattacharyya:2018wym,Caputa:2020mgb,Yang:2020tna}.
 
 One of the main advantages of the path integral optimization procedure as well as path integral complexity is that one can employ standard holographic dictionary to find their gravity dual. Indeed, as it was recently explained in~\cite{Boruch:2020wax,Boruch:2021hqs} (see review below), preparing a state with a Euclidean path integral on a curved geometry in a holographic CFT can be described as computing the Hartle-Hawking wavefunction up to some bulk slice $B$ with an arbitrary induced metric $h$. The maximization of the gravity wavefunction with respect to this metric yields the same geometry as the boundary path integral optimization.  Moreover, the Einstein-Hilbert action with appropriate Hayward terms evaluated from the boundary up to surface $B$ can be seen as the ``full" path integral complexity action: i.e. the CFT path-integral complexity action (i.e., Liouville action in $2$d) with finite cut-off contributions. This new gravity perspective not only allows to derive the Liouville action from the bulk but also gives a prediction for its higher-dimensional as well as Lorentzian generalisations \cite{Boruch:2021hqs,Caputa:2021pad}. Interestingly, the ultraviolet (UV) limit of the holographic path integral complexity reproduces complexity actions proposed for higher-dimensional CFTs in \cite{Caputa:2017yrh,Caputa:2017urj} that are all two-derivative i.e., first order in the curvature of the path integral background. Nevertheless, these higher-dimensional complexity actions and their operational interpretation are much less explored and understood than their $2$d counterpart.
 
 These holographic results bring new questions to the path integral complexity proposal. Firstly, especially in higher dimensions, one would like to better understand the relation between the ``full" optimization actions from gravity and their UV limits. For example, one may wonder how to systematically include finite cut-off (curvature) corrections to the present CFT optimization actions. If one wanted to repeat the arguments of the derivation of complexity action from the CFT wavefunctions, at least in even dimensions, one would be naturally lead to anomaly actions that are higher curvature in the background metric. They can be systematically organised into Q-curvature actions and we will discuss this approach in more details below. On the other hand, one could try to understand the UV limit of the full gravity actions in terms of penalty factors. In the approach of Nielsen~\cite{Nielsen1133}, penalty factors are arbitrary functions which are meant to control how much a particular operator contributes to the depth of the circuit, thus providing a way of distinguishing between gates which are ``easy" to implement, and gates which are ``hard".  An outstanding open question in this regard is the role that such penalty factors play in holography (see also recent discussion \cite{Erdmenger:2021wzc}), in particular in the path integral optimization as well as its gravitational interpretation. Secondly, from the gravity side, one may wonder whether there is a natural way of  modifying the holographic path integral optimization such that one could ``tune" the higher derivative terms in the UV limit. This question is closely related to the above-mentioned finite cut-off corrections to the boundary complexity action and their interpretation.

 In this work we make a modest progress on these questions. We start by formulating the path integral optimization as a uniformization problem in even dimensions, which resorts to Q-curvature actions~\cite{Levy:2018bdc, Chernicoff:2018apt}. The Q-curvature actions can be considered systematically as the higher-dimensional generalizations of the Liouville action and their optimization also provides the hyperbolic geometries as saddles. We explore their properties (such as e.g., co-cycle conditions), differences with two-derivative complexity actions \cite{Caputa:2017yrh,Caputa:2017urj} and their TN interpretations. In the second part we consider adding higher curvature terms on the surface $B$ in the holographic computation of the Hartle-Hawking wavefunctions and discuss its consistency with the finite cut-off holography and $T^2$-deformations of holographic CFTs.

 The paper is structured as follows: we first briefly review the optimization of Euclidean path integrals and its holographic interpretation in section~\ref{sec:pathint}, discussing the higher-dimensional complexity action and the need for a better understanding from the perspective of CFTs. We then formulate the path integral optimization as an uniformization problem in terms of the Q-curvature in section~\ref{sec:unif}. We discuss some solutions of the constant Q-curvature constraint as well as the TN interpretation of the uniformization problem providing an interpretation of the penalty factors from this perspective. In section~\ref{sec:gra}, we discuss the effect of adding higher-curvature terms to the brane action in the Hartle--Hawking wavefunction approach and verify the consistency of the optimization with another approach to holographic tensor networks based on the $T\bar{T}$ deformations \cite{Caputa:2020fbc}. 	
 
\section{ Path Integral Optimization in CFTs and Holography} \label{sec:pathint}

We start by briefly reviewing the path integral optimization in CFTs~\cite{Caputa:2017urj,Caputa:2017yrh} and its holographic interpretation~\cite{Boruch:2020wax,Boruch:2021hqs}. Most of this material is described pedagogically in original works so readers should consult them for further details and clarifications.
	
The goal of the \emph{path integral optimization}~\cite{Caputa:2017urj,Caputa:2017yrh} is to sharpen the intuitions behind the emergence of co-dimension-one slices of holographic geometries from TN in CFTs~\cite{Swingle:2012wq} (see also \cite{Miyaji:2016mxg,Bao:2018pvs,Caputa:2020fbc,Chen:2021ipv}). The main object of interest, for a CFT defined in $d$-dimensional flat Euclidean spacetime $\mathbb{R}^{d}$, is the Euclidean path integral that prepares a ground state
\begin{align} \label{eq:GroundCFTPathInt}
\begin{split}
&\Psi_{0}[\tilde{\varphi}(\vec{x})]:=\lim_{\beta\rightarrow \infty}\braket{\tilde{\varphi}(\vec{x})\vert e^{-\beta\, \hat{H}_{\textrm{CFT}}}\vert\Psi_{0}} = \int [ \mathcal{D}\varphi]\,e^{-I_{\textrm{CFT}}[\varphi]}\,\delta\left(\varphi\vert_{\partial(\mathbb{R}^{d})}-\tilde{\varphi}\right)\\
&=\int\left(\prod_{\vec{x}}\,\prod_{\epsilon\leq z<\infty}\,\mathcal{D}\varphi(z,\vec{x})\right)e^{-I_{\textrm{CFT}}[\varphi]}\times\prod_{\vec{x}}\delta\left(\varphi(\epsilon,\vec{x})-\tilde{\varphi}(\vec{x})\right)~,
\end{split}
\end{align}
where $\epsilon$ is a UV cut-off identified with a lattice spacing in a discretized setting, $(z,\vec{x})$ are local coordinates in $\mathbb{R}^{d}$, $\tau:=-z$ is the Euclidean time and $\vec{x}=(x^{1},\ldots,x^{d-1})$ are local coordinates in $(d-1)$-dimensional Euclidean space $\mathbb{R}^{d-1}$. $I_{\textrm{CFT}}$ is the CFT action given in terms of the fields $\varphi(z,\vec{x})$, whose boundary condition at $z=\epsilon$ is $\tilde{\varphi}(\vec{x})$.

\begin{figure}[tb]
\centering
\includegraphics[width=0.8\textwidth]{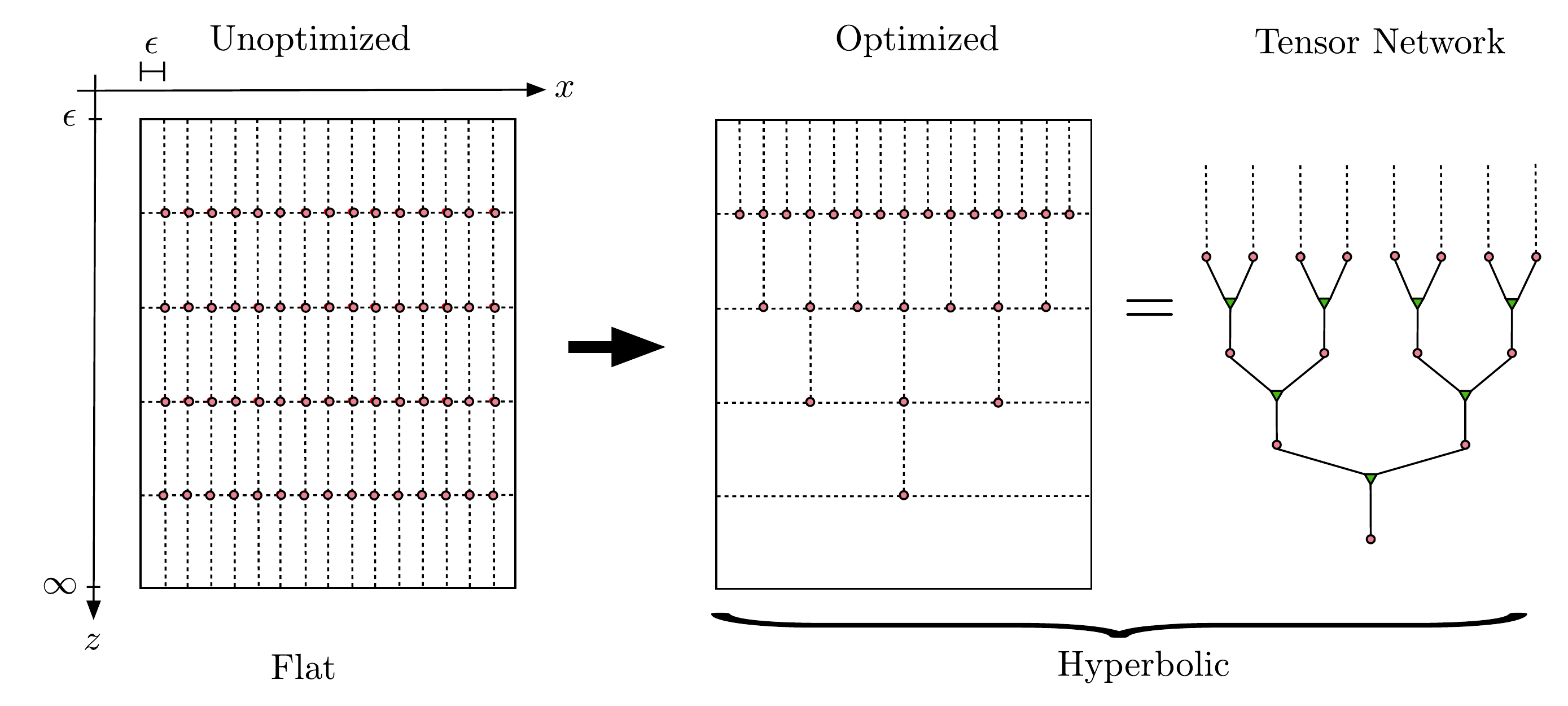}
\caption[Representation of the optimization procedure of Euclidean path integrals.]{Representation of the optimization procedure of Euclidean path integrals. The small pink circles represent lattice sites. On the left side, the computation of the ground state wavefunctional from the Euclidean path integral~\eqref{eq:GroundCFTPathInt} on a flat unoptimized background. On the middle, the optimized path integral described by a hyperbolic geometry. On the right, a tensor network representation of the optimized hyperbolic geometry where only the coarse-graining procedure of lattice sites is shown.}
\label{fig:PathIntOpt}
\end{figure}

We then perform an ``optimization" of~\eqref{eq:GroundCFTPathInt}, which can be intuitively visualised in the following way: we first discretize $\mathbb{R}^{d}$ into a square, evenly-spaced ``unoptimized" lattice, as shown in the left panel of Fig.~\ref{fig:PathIntOpt}.  Next, we optimize this lattice by effectively removing the unnecessary lattice sites on which the path integral is computed. This can be interpreted as a ``coarse-graining" procedure where only low-energy modes $\vert \vec{k} \vert\ll -1/z=1/\tau$ remain in the path integral for a given time $\tau$. This implies that a number of lattice sites of order $\mathcal{O}(\tau/\epsilon)$ can be combined into one without losing much accuracy in the evaluation of~\eqref{eq:GroundCFTPathInt}. This optimization procedure of the path integral is represented in the middle panel of Fig.~\ref{fig:PathIntOpt}. The optimized lattice can be interpreted in the continuum as hyperbolic metric (TN) over which the Euclidean path integral computes the CFT ground state $\ket{\Psi_{0}}$. 

The above intuition can be implemented directly in the continuum. Namely, we start with a metric on the $d$-dimensional flat space $\mathbb{R}^{d}$ in such a way that there is a single lattice site per unit area. The unoptimized metric can be written in local coordinates $(z,\vec{x})$ as
\begin{align} \label{eq:FlatMetric}
\textrm{d}s^{2}=\frac{1}{\epsilon^{2}}\left(\textrm{d}z^{2}+\textrm{d}\vec{x}^{2}\right)~.
\end{align}
The optimization is then performed at the level of the metric, by replacing it with a general metric of the form
\begin{align} \label{eq:GenMetricTBO}
\textrm{d}s^{2}=g_{zz}(z,\vec{x})\textrm{d}z^{2}+\sum_{i,j=1}^{d-1}g_{ij}(z,\vec{x})\textrm{d}x^{i}\,\textrm{d}x^{j}+2\sum_{j=1}^{d-1}g_{z,j}(z,\vec{x})\textrm{d}z\,\textrm{d}x^{j}~,
\end{align}
subject to the constraint that \eqref{eq:FlatMetric} is reproduced at $z=\epsilon$, so that the UV regularization for the optimized metric agrees with the original one~\eqref{eq:FlatMetric} at the end of the path-integration. Moreover, the boundary conditions for all the fields are held fixed in this procedure so that the wavefunctions in these two geometries are proportional to each other.\footnote{A crucial observation is that in CFTs it is possible to perform the optimization of~\eqref{eq:GroundCFTPathInt} by only changing the background metric as in~\eqref{eq:GenMetricTBO}. This not the case for non-conformal theories or for CFTs in the presence of external fields since in this case said fields, such as the mass or other couplings, must be modified in a position dependent way due to the coupling's renormalization group (RG) flow.}
For example, in $2$-dimensions, all metrics $g_{ab}$ can be brought to a conformally-flat diagonal form via a coordinate transformation. Therefore, without any loss of generality we can consider a $2$-dimensional metric written in local coordinates $(z,x)$ as
\begin{align} \label{eq:2dConfMetric}
\textrm{d}s^{2}=e^{2\phi(z,x)}\left(\textrm{d}z^{2}+\textrm{d}x^{2}\right)~,
\end{align}
where the \emph{Weyl} (or conformal) \emph{factor} $\phi(z,x)$, which contains all the information about the metric, is subject to the following boundary condition\footnote{Here we are following the conventions used in \cite{Caputa:2017urj, Caputa:2017yrh}. To be dimensionally accurate, we need to restore $\epsilon$ in appropriate places. For example, in the case the boundary condition \eqref{eq:2dBdryCond} would be simply $e^{2\phi(z=\epsilon,x)} = 1$.}
\begin{align} \label{eq:2dBdryCond}
e^{2\phi(z=\epsilon,x)}=\frac{1}{\epsilon^{2}}=e^{2\phi_{0}(x)}~.
\end{align}
Moreover, it is well known~\cite{Polyakov:1981rd} that while Weyl rescaling is a symmetry of the CFT action, it leads to anomalous transformation of the path integral measure such that 
\begin{align} \label{eq:WaveFuncResc}
\Psi[\tilde{\varphi}(\vec{x})]\Big\vert_{g_{ab}=e^{2\phi}\delta_{ab}}=e^{I_{L}[\phi]-I_{L}[0]}\cdot \Psi[\tilde{\varphi}(\vec{x})]\Big\vert_{g_{ab}=\delta_{ab}}~,
\end{align}
where $I_{L}[\phi]$ is the famous \emph{Liouville action}
\begin{align}\label{eq:LiouvilleAct}
I_{L}[\phi]=\frac{c}{24\pi}\int^{+\infty}_{-\infty}\textrm{d}x\int^{+\infty}_{\epsilon}\textrm{d}z\,\left((\partial_{x}\phi)^{2}+(\partial_{z}\phi)^{2}+\mu\,e^{2\phi}\right)~,
\end{align}
where $c$ is the central charge of the CFT and where $\mu$ is an $\mathcal{O}(1)$ constant identified with $1/\epsilon^{2}$ in a discretized setting. The kinetic term in the Liouville action~\eqref{eq:LiouvilleAct} is proportional to the Ricci scalar and describes the conformal anomaly in two dimensions, while the potential term $\mu e^{2\phi}$ arises from the UV regularization. As such, the potential term should dominate over the kinetic term as the UV cut-off $\mu\sim1/\epsilon^{2}$ is taken to infinity, which is realized when
\begin{align}\label{eq:RelKinetPoten}
(\partial_{i}\phi)^{2}\ll e^{2\phi}\,\,,\,\,(i=z,x)~.
\end{align}
Given this observation, it was proposed in~\cite{Caputa:2017urj} that the optimization of the path integral should be done by choosing the background metric that minimizes the Liouville action subjected to boundary conditions \eqref{eq:2dBdryCond}.
In other words, optimal metrics should solve the \emph{Liouville equation} which is in fact equivalent to the constraint that the Ricci scalar $R$ of the $2$-dimensional metric~\eqref{eq:2dConfMetric} is constant
\begin{align} \label{eq:LiouvilleEq}
(\partial_{x}^{2}+\partial_{z}^{2})\phi=\mu e^{2\phi}\qquad \Leftrightarrow\qquad R=-2\mu.
\end{align}
A solution to this equation which satisfies the boundary condition~\eqref{eq:2dBdryCond} is given by the Weyl factor and metric of the hyperbolic plane
\begin{align}\label{eq:PoincDiscH2}
e^{2\phi}=\frac{1}{\mu z^{2}},\qquad \textrm{d}s^{2}=\frac{1}{\mu z^{2}}(\textrm{d}z^{2}+\textrm{d}x^{2})~.
\end{align}
This hyperbolic metric with $\mu=1$ corresponds in fact to the minimum of the Liouville action~\eqref{eq:LiouvilleAct} satisfying the boundary condition~\eqref{eq:2dBdryCond} as can be seen by rewriting the former as
\begin{align} \label{eq:HyperbMinimumLiouville}
I_{L}=\frac{c}{24\pi}\int \,\textrm{d}x\,\textrm{d}z\left[(\partial_{x}\phi)^{2}+(\partial_{z}\phi+ e^{\phi})^{2}\right]-\frac{c}{12\pi}\int \,\textrm{d}x\left[ e^{\phi}\right]^{z=\infty}_{z=\epsilon}\geq \frac{c\,L_{x}}{12\pi\epsilon}~,
\end{align}
where $L_{x}=\int\,\textrm{d}x$ is the infinite volume (length in this case) of the spatial $x$ direction. However,  as we will discuss momentarily, one can view the metrics arising from~\eqref{eq:PoincDiscH2} as corresponding to different degrees of optimization for different values of $0<\mu\leq 1$ with $\mu=1$ corresponding to the maximally optimized geometry.

The appearance of the hyperbolic space from the optimization was interpreted as an explicit realization of the AdS/TN correspondence in which such TN could be thought of as a slice of the holographic AdS$_{3}$. In~\cite{Caputa:2017urj,Caputa:2017yrh} it was also shown that the geometries obtained via the optimization of Euclidean path integrals for other states in $2$-dimensional CFTs such as excited states (given by primaries) or thermal states lead consistently to time-slices of AdS$_{3}$ and the proposal for general spacetimes was described in \cite{Takayanagi:2018pml}.

However, a subtle issue arises when taking a closer look at the hyperbolic solution~\eqref{eq:PoincDiscH2}. In this case, $(\partial_{i}\phi)^{2}$ and $e^{2\phi}$ are found to be of the same order, which is at odds with the expectation~\eqref{eq:RelKinetPoten} obtained in the limit where the UV cut-off $\epsilon$ is taken to infinity. This observation suggests that the path integral optimization via the Liouville action is in fact qualitative and therefore there should be finite cut-off corrections to this procedure. For example in the explicit Heat-Kernel derivation of \eqref{eq:WaveFuncResc} for free theories, one neglects higher curvature terms that are suppressed with powers of the UV cut-off. The main open question is how such terms should be included and under what assumptions (e.g. holographic CFTs) this can be done universally.

\subsection{Path Integral Complexity}
\label{subsec:PIOComp}
Intuitively, the optimization of the path integral that prepares a wavefunction corresponds to a minimization of the number of operations that need to be performed in the discretized description. This discrete Euclidean path-integration can be then mapped into a TN, whose optimization can be carried out by \emph{tensor network renormalization} (TNR)~\cite{Evenbly_2015}. In this sense, the optimization of Euclidean path integrals is a natural counterpart of TNR. This implies an interesting connection between the optimization and a notion of \emph{complexity}, as measured by the number of tensors that are needed to construct the TN. Indeed, one can intuitively associate a notion of complexity to a state represented by a TN by counting the number of tensors (volume of the optimal TN) that are needed to accurately represent it: the more tensors are needed, the more ``complex" the state is. 

This naturally led to a notion of \emph{path integral complexity}  as described in~\cite{Caputa:2017yrh}, where the complexity $\mathcal{C}_{\Psi}$ of a CFT state $\ket{\Psi}$ is obtained by minimizing the functional $I_{\Psi}[g_{ab}(z,\vec{x})]$ defined by the ratio of the two wavefunctions
\begin{align}\label{RationofWF}
 I_{\Psi}[g_{ab}(z,\vec{x})]\equiv \log\left(\frac{\Psi_{g_{ab}}}{\Psi_{\delta_{ab}}}\right),
\end{align}
and the actual complexity of $\vert\Psi\rangle$ is given by the on-shell value
\begin{align}\label{eq:PathIntComp}
\mathcal{C}_{\Psi}:=\min_{g_{ab}(z,\vec{x})}[I_{\Psi}[g_{ab}(z,\vec{x})]]~.
\end{align}
That is, the functional $I_{\Psi}[g_{ab}(z,\vec{x})]$ estimates the complexity of the TN corresponding to the path integral computed for a specific metric $g_{ab}$ relatively to $g_{ab}=\delta_{ab}$.

This path integral complexity~\eqref{eq:PathIntComp} acquires a precise realization in the case of $2$-dimensional CFTs given the identification of the functional which determines the path integral optimization with the Liouville action~\eqref{eq:LiouvilleAct}. In particular, since the hyperbolic geometry~\eqref{eq:PoincDiscH2} saturates the bound~\eqref{eq:HyperbMinimumLiouville}, this means that the path integral complexity for the ground state of $2$-dimensional CFTs is given by the Liouville action on the hyperbolic geometry and is also proportional to the spatial volume
\begin{align}\label{eq:PathIntCOmp2D}
\mathcal{C}_{\Psi_{0}}=\min_{\phi}[I_{L}[\phi]]=\frac{c\,L_{x}}{12\pi\epsilon}~,
\end{align}
a result which agrees with the expected leading UV behaviour of the ground state of a CFT.

This connection between the Liouville action and a notion of complexity in $2$-dimensional CFTs through path integral optimization has been further generalized to various CFTs and QFTs (see \emph{e.g.}~\cite{Bhattacharyya:2018wym,Jafari:2019qns,Ghodrati:2019bzz,Caputa:2020mgb,Ahmadain:2020jfm}), and has also been connected with more direct approaches to circuit complexity~\cite{Caputa:2018kdj,Camargo:2019isp}. Moreover, in connection with the TN interpretation of complexity, it was proposed in~\cite{Czech:2017ryf} that the terms appearing in the Liouville action~\eqref{eq:LiouvilleAct} correspond to tensors in MERA. Qualitatively, the kinetic terms $(\partial_{x}\phi)^{2}+(\partial_{z}\phi)^{2}$ corresponding to isometries and the potential term $e^{2\phi}$ to unitaries. Similarly, authors in~\cite{Camargo:2019isp} discussed a relation between the path integral complexity measured by the Liouville action~\eqref{eq:LiouvilleAct} and a notion of circuit complexity arising from non-unitary circuits built from components of the stress tensor in $2$-dimensional CFTs \cite{Caputa:2018kdj,Milsted:2018vop}. In particular, they observed that one way of extending the Liouville action to finite cut-off corrections could be done by considering a complexity functional (cost function) resembling the well known \emph{Dirac--Born--Infeld} (DBI) action~\cite{Polchinski:1996na}\footnote{See also \cite{Jatkar:2011ue} for such structure in the holographic counter-term actions.} 
\begin{align} \label{eq:DBIAction}
I_{\textrm{DBI}}\propto-\tilde{T}\int \textrm{d}^{2}\chi(z,x)\sqrt{-\textrm{det}\left(g_{ab}+\epsilon^{2}\partial_{a}\chi(z,x)\partial_{b}\chi(z,x)\right)},~
\end{align}
where $\tilde{T}$ is known as the \emph{brane tension}, which is proportional to $\mathcal{O}\left((G^{(3)}_{N})^{-1}\right)\propto c$, and where $\chi(x,z)=(\chi_{1}(x,z),\chi_{2}(x,z))$ represents a coordinate transformation from the original flat coordinates $(z,x)$ to curvilinear coordinates $(\chi_{1},\chi_{2})$. Even though this guess was not derived in any systematic way from CFTs in \cite{Camargo:2019isp}, we will see below that complexity actions arising from gravity optimization indeed hint on similar structures.

\subsection{Holographic Path Integral Optimization}
\label{sec:PIOHH}
\begin{figure}[tb]
\centering
\includegraphics[width=0.65\textwidth]{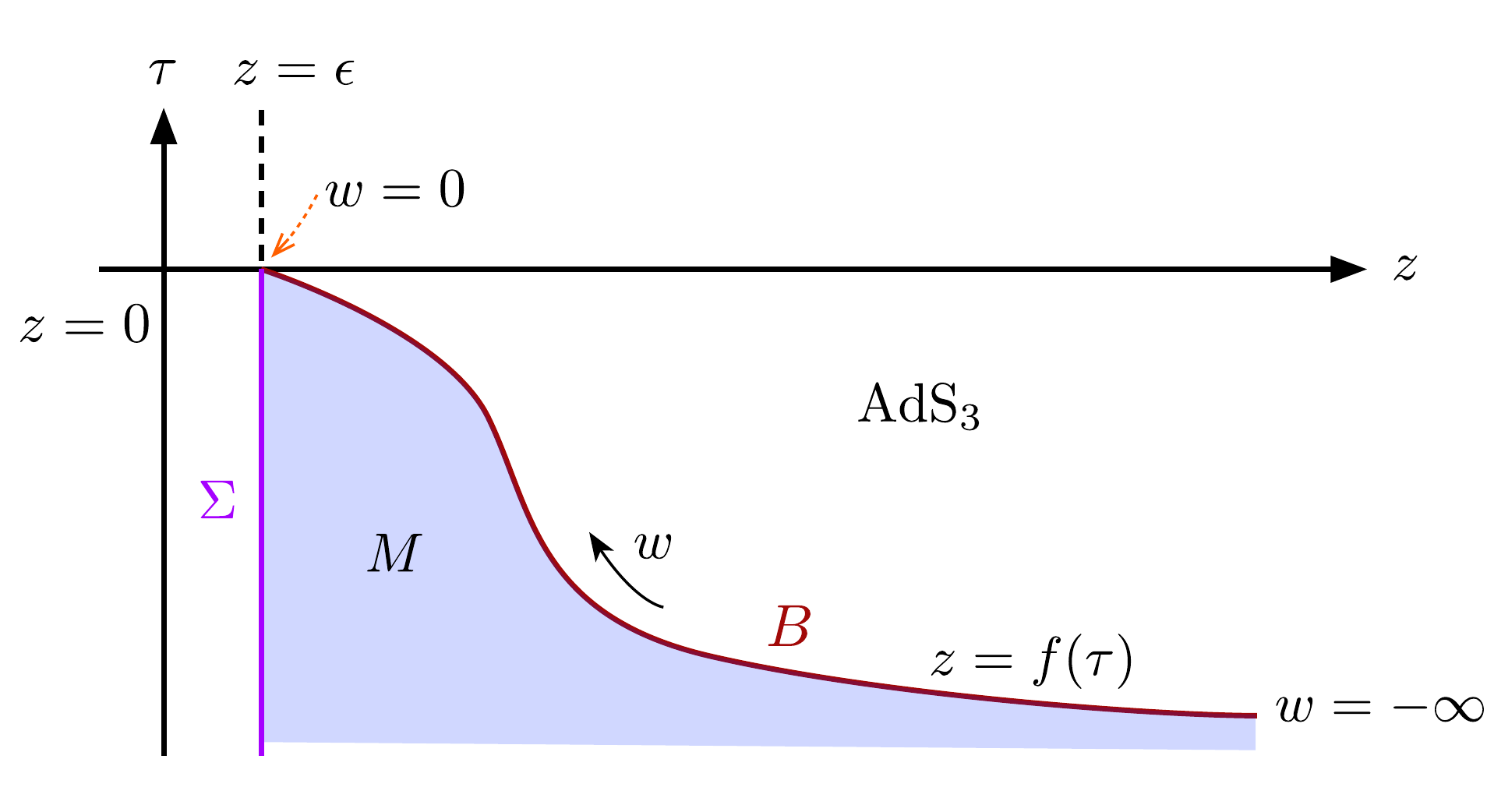}
\caption[Diagram of the geometric region over which an evaluation of the gravitational action yields the computation of the Hartle--Hawking wavefunctional in AdS$_{3}$.]{Diagram of the geometric region $M$ over which an evaluation of the gravitational action $I_G$ yields the computation of the Hartle--Hawking wavefunctional in AdS$_{3}$.}
\label{fig:HHGeometry}
\end{figure}

As mentioned above, a recent proposal~\cite{Boruch:2020wax,Boruch:2021hqs} provides a dual description of the path integral optimization procedure from the gravitational perspective within the AdS/CFT correspondence in terms of the \emph{Hartle--Hawking wavefunctional}~\cite{Hartle:1983ai} taken to evolve from the boundary of AdS up to a certain slice of the bulk. This corresponds to an evaluation of the gravitational action in the blue shaded region in Fig.~\ref{fig:HHGeometry}, computed for an Euclidean AdS$_{d+1}$ geometry written in Poincar\'{e} coordinates $(z,\tau,x^{i})$
\begin{align} \label{eq:EucAdSPoinc}
\textrm{d}s^{2}=\frac{1}{z^{2}}\left(\textrm{d}z^{2}+\textrm{d}\tau^{2}+\textrm{d}\vec{x}^{2}\right)~.
\end{align}
More precisely, the idea is to consider the Hartle--Hawking (HH) wavefunctional $\Psi_{\textrm{HH}}[g_{ab}]$ in an asymptotically AdS$_{d+1}$ spacetime which evaluates the path integral of Euclidean gravity from a cut-off surface $\Sigma$ near the asymptotic boundary given by $z=\epsilon$ and $\tau<0$ to the surface $B$, given by $z=f(\tau)$, which is located in the bulk and stems from $z=\epsilon$ and $\tau=0$. See Fig.~\ref{fig:HHGeometry}. The HH wavefunctional is defined as
\begin{align} \label{eq:HHWaveDef}
\Psi_{\textrm{HH}}[g_{ab}]:=\int [\mathcal{D}g_{ab}]e^{-I_{\textrm{G}}[g_{ab}]}\delta(g_{ab}\vert_{B}-e^{2\phi}\delta_{ab})~,
\end{align}
where the metric on the surface $B$ is assumed to have the translational invariant form
\begin{align} \label{eq:QsurfMetric}
\textrm{d}s^{2}=e^{2\phi}(\textrm{d}w^{2}+\textrm{d}\vec{x}^{2})~,
\end{align}
where the Weyl factor $\phi(w,\vec{x})$ contains all the relevant information about the metric~\eqref{eq:QsurfMetric}.\footnote{One can also take more general metric on $B$ but it would require starting from a more complicated solution of Einstein's equations.} One should note that this procedure contemplates a semiclassical computation of the path integral~\eqref{eq:HHWaveDef}. Another remark is that there is an implicit dependence of the coordinate $w$ which characterizes the surface $B$ and the Euclidean time $\tau$ defined on the AdS space: $w=w(\tau)$. 

The gravitational action $I_{\textrm{G}}$ on the $(d+1)$-dimensional AdS spacetime which contains a bulk and Gibbons--Hawking--York (GHY) boundary contributions is given by
\begin{align} \label{eq:GravAcHH}
I_{\textrm{G}}=-\frac{1}{16\pi G^{(d+1)}_{N}}\int_{M}\textrm{d}^{d+1}x\sqrt{g}\left(R-2\Lambda\right)-\frac{1}{8\pi G^{(d+1)}_{N}}\int_{B\cup \Sigma}\textrm{d}^{d}x\sqrt{h}\,K~,
\end{align}
where $\Lambda$ is the cosmological constant, $R$ is the Ricci scalar of the $(d+1)$-dimensional AdS spacetime, $g$ is the determinant of the metric~\eqref{eq:EucAdSPoinc}, $h$ is the determinant of the induced metric on $B\cup \Sigma$ and $K$ is the trace of the extrinsic curvature also on $B\cup \Sigma$.

Another crucial ingredient to this interpretation is that the surface $B$ in the bulk should be looked at as a \emph{probe brane} which extends from the boundary $\Sigma$ and into the bulk, according the AdS/BCFT~\cite{Takayanagi:2011zk,Fujita:2011fp} prescription. That is, one adds a tension term on $B$ to~\eqref{eq:GravAcHH} given by
\begin{align} \label{eq:BraneTension}
I_{\textrm{T}}=\frac{T}{8\pi G^{(d+1)}_{N}}\int_{B} \textrm{d}^{d}x\,\sqrt{h}~,
\end{align}
which is proportional to the volume of the surface $B$ and whose contribution to the gravitational action~\eqref{eq:GravAcHH} is controlled by the sign of the \emph{tension} $T$. In such a way, one obtains a one-parameter family of deformed HH wavefunctionals given by
\begin{align} \label{eq:HHWaveBraneDef}
\Psi_{\textrm{HH}}^{(T)}[\phi]:=\int [\mathcal{D}g_{ab}]e^{-I_{\textrm{G}}[\phi]-I_{\textrm{T}}[e^{2\phi}]}\delta(g_{ab}\vert_{B}-e^{2\phi}\epsilon_{ab})~,
\end{align}
from which the standard HH wavefunctional~\eqref{eq:HHWaveDef} is obtained by setting $T=0$. Note that it is also important that the brane $B$ does not back-react on the AdS geometry.

\begin{figure}[tb]
\centering
\includegraphics[width=0.4\textwidth]{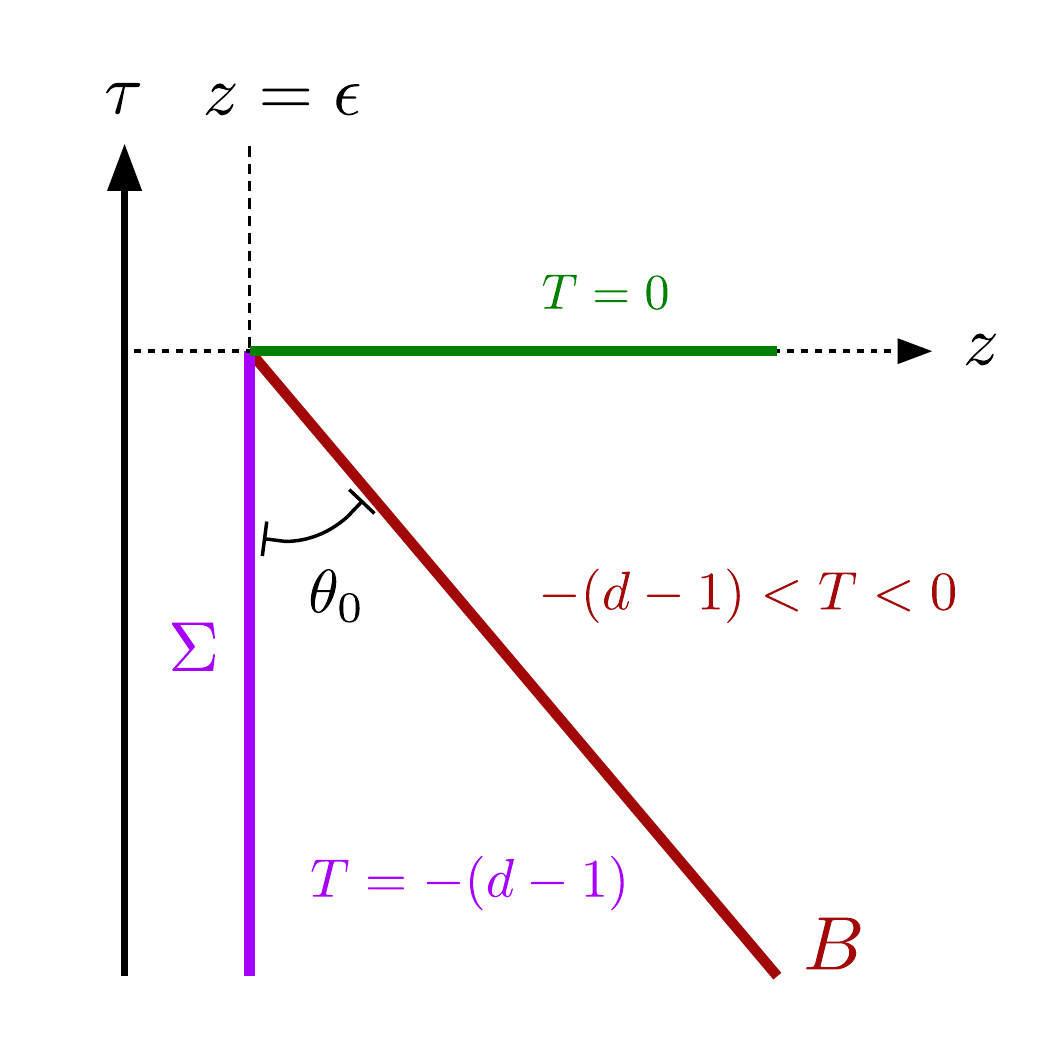}
\caption[Diagram of the solution which maximizes the deformed Hartle--Hawking wavefunction with an end-of-the-world brane in AdS$_{3}$.]{The brane $B$ interpolates between the boundary $\Sigma$ at $T=-(d-1)$ and the $\tau=0$ time slice at $T=0$. The angle $\theta_{0}$ between $B$ and $\Sigma$ is given by $\theta_{0}=\arcsin\left(1-T^{2}/(d-1)^{2}\right)^{1/2}$.}
\label{fig:HHSolutionT}
\end{figure}

These deformed HH wavefunctionals can be evaluated semi-classically using the saddle-point approximation. In particular, the actions $I_{\textrm{G}}+I_{\textrm{T}}$ can be evaluated directly and, for example in $2$ dimensions, neglecting finite cut-off corrections and assuming $(\partial_{i}\phi)^{2}\ll e^{2\phi}$ one reproduces the Liouville action together with the optimal geometries derived for various universal classes of CFT states. For example, surfaces $B$ for the vacuum state are given by half-planes (see Fig. \ref{fig:HHSolutionT})
\begin{align} \label{eq:SolHHWaveMax}
z=\epsilon+\tau\frac{\sqrt{1-T^2}}{T}~,
\end{align}
parametrized by $-1<T<0$ and their induced metric matches the 2d surface from the Liouville optimization for the vacuum. In particular, the coefficient $\mu$ in the Liouville action translates into the tension parameter
\begin{align} \label{eq:RelMuTension}
\mu=1-T^{2}~.
\end{align}
 As we saw previously, the parameter $\mu$ can be thought of as measuring how optimized the background metric (TN) is within the path integral optimization scheme. As a consequence, from the gravitational perspective this corresponds to changing the tension $T$ from $-1$ to $0$, where $T=0$ corresponds to fully-optimized solution. Geometrically, this variation of the tension positions the boundary-anchored brane $B$ moving from the boundary $\Sigma$ at $T=-1$ to a time slice $\tau=0$, as can be seen in Fig.~\ref{fig:HHSolutionT}.

In general dimensions $d$, varying the on-shell action $I_{\textrm{G}}+I_{\textrm{T}}$ is equivalent to imposing the Neumann boundary condition on $B$, consistent with the AdS/BCFT construction, given by
\begin{align} \label{eq:NeuBoundCond}
K_{ab}-Kh_{ab}=-Th_{ab}~,
\end{align}
where $K_{ab}$, $K$ and $h_{ab}$ are respectively the extrinsic curvature, its trace and the induced metric on $B$. Note that by the Hamiltonian constraint, which is always satisfied for on-shell solutions, this implies
\begin{align} \label{eq:HamConst}
K^{2}-K^{ab}K_{ab}=\frac{d}{d-1}T^{2}=R-2\Lambda~,
\end{align}
where $R$ is the Ricci scalar on $B$ and $\Lambda$ is the cosmological constant of AdS$_{d+1}$, and where we substituted $K\vert_{B}= T\, d/(d-1)$ that is just the trace of $\eqref{eq:NeuBoundCond}$. This is another confirmation of the holographic path integral optimization since, after inserting \eqref{eq:RelMuTension}, this constraint becomes precisely the CFT optimization \eqref{eq:LiouvilleEq}.

While the maximization of the HH wavefunctional can be performed unambiguously for any dimension $d$, and gives a clear prediction for the CFT path integral complexity action in the UV limit, there are still important questions regarding the precise optimization procedure in higher-dimensional CFTs.  

\subsection{Higher-Dimensional CFTs}
\label{subsec:PIOhdCFTOrig}
A natural question in the context of path integral optimization is whether an explicit form of the functional $I_{\Psi}[g_{ab}]$~\eqref{eq:PathIntComp} whose minimization leads to the optimization of the Euclidean path integral $\Psi_{0}[\tilde{\varphi}(\vec{x})]$ can be found in higher dimensions. This is also necessary to determine the path integral complexity $\mathcal{C}_{\Psi}$ in higher-dimensional CFTs. On this matter, there exists a proposal for ``effective" path integral complexity action~\cite{Caputa:2017urj,Caputa:2017yrh} $I_{\Psi}[g_{ab}]$ constructed in the following way: Starting from a metric $g_{ab}$ of the form
\begin{align}\label{eq:HigherDPathIntMetric}
\textrm{d}s^{2}=g_{ab}\textrm{d}x^{a}\textrm{d}x^{b}=e^{2\phi(x)}\hat{g}_{ab}\textrm{d}x^{a}\textrm{d}x^{b}~,
\end{align}
the following action should be minimized for a vacuum state of a $d$-dimensional CFT (as well as some small excitations around the ground state)
\begin{align}\label{eq:HigherDPathIntAction}
I_{\Psi}[\phi,\hat{g}]:=\frac{d-1}{16\pi G^{(d)}_{N}}\int\textrm{d}^{d}x\sqrt{\hat{g}}\left(e^{(d-2)\phi}\hat{g}^{ab}\partial_{a}\phi\partial_{b}\phi+\frac{e^{(d-2)\phi}R_{\hat{g}}}{(d-1)(d-2)}+\mu e^{d\phi}\right)~,
\end{align}
where $R_{\hat{g}}$ is the Ricci scalar of the metric $\hat{g}_{ab}$. Among various other features which led to this identification is the fact that such a functional satisfies the so-called co-cycle conditions \cite{Caputa:2017urj,Caputa:2017yrh}. Interestingly, \eqref{eq:HigherDPathIntAction} can be re-written as the Einstein--Hilbert action in $d$-dimensions with negative cosmological constant $\Lambda^{(d)}=-(d-1)(d-2)/2$. This generalized the optimization equation obtained by variation with respect to $\phi(x)$ that implies taking the trace of vacuum Einstein's equations, i.e., the condition that the Ricci scalar of \eqref{eq:HigherDPathIntMetric} should be a negative constant. Last but not the least, the action \eqref{eq:HigherDPathIntAction} was also reproduced in the UV limit of the holographic path integral complexity action~\cite{Boruch:2021hqs} explained in the previous section.

Despite these non-trivial consistency checks and observations, there are  still some puzzles when identifying the functional~\eqref{eq:HigherDPathIntAction} as a higher-dimensional generalization of the Liouville action. Firstly, from the perspective of the action itself it is not clear why it should be restricted to having quadratic derivatives of the Weyl field $\phi$. Generally, it is quite natural in AdS/CFT that (``sub-leading") higher-derivative terms will also contribute in higher dimensions. This is similar to the problem of the gravitational action in spacetime dimensions higher than $d+1=4$ in which one generically views the Einstein--Hilbert action as a low energy effective theory containing only terms that are quadratic in the derivatives of the metric.\footnote{This is best seen by considering Lovelock's theorem~\cite{Lovelock:1971yv,Lovelock:1972vz} which is used to construct natural higher-dimensional generalizations of Einstein gravity which include higher-curvature corrections. These so-called \emph{Lovelock theories} are metric theories of gravity which lead to conserved second order equations of motion that naturally take into account higher-curvature terms in the action which become topological in lower-dimensional theories.} 

This is even more pronounced once we consider even-dimensional CFTs and intend to define the complexity functional from the ratio of wave functions \eqref{RationofWF}. This would naturally lead to the so-called anomaly actions of the Riegert type~\cite{Riegert:1984kt} that are also referred to as Q-curvature actions~\cite{Levy:2018bdc, Cercle:2019jxx,Freidel:2008sh}.  For example, in 4d  holographic CFTs with central charges $a=c$ the Weyl anomaly reads
\begin{equation}
\langle T^\mu_{\phantom{\mu}\mu}\rangle =\frac{c}{2\pi} \mathcal{Q}_4,
\end{equation}
and is responsible for the transformation of partition functions (see e.g. \cite{Freidel:2008sh})
\begin{equation}\label{QCAPF}
Z_{\textrm{CFT}}(e^{2\phi}\hat{g})=e^{\frac{c}{4\pi}\int\sqrt{\hat{g}}\left(\phi\mathcal{P}_4\phi+2 \mathcal{Q}_4\phi\right) }Z_{\textrm{CFT}}(\hat{g}),
\end{equation}
where the Q-curvature $\mathcal{Q}_4$ and $\mathcal{P}_4$ will be discussed below. There is a similar expectation in 6d  holographic CFTs, where the six-dimensional Q-curvature $\mathcal{Q}_6$ captures the type-A anomaly directly related to the six-dimensional Euler density $E_{6}$ (see~\cite{Henningson:1998gx, Beccaria:2015ypa, Bugini:2016nvn, Lu:2019urr} for more details).

Similarly as in 2d, we may expect that the action~\eqref{QCAPF} will play a similar role to Liouville in the optimization of the holographic (at least those with holographic Weyl anomalies \cite{Henningson:1998gx}) CFT wavefunctions. In the following sections, we will follow this CFT prediction, and discuss similarities and differences between higher-dimensional path-integral optimization done with the Q-curvature actions as in \eqref{QCAPF} and \eqref{eq:HigherDPathIntAction} proposed in \cite{Caputa:2017urj,Caputa:2017yrh}.

Last but not the least, from the gravitational perspective it is an interesting question how other geometrical or physical (\emph{e.g.} matter) properties of the surface $B$ could be incorporated in the holographic path integral proposal. In a precise sense, the surface $B$ can be understood as a time-dependent cut-off \cite{Caputa:2020fbc,Chandra:2021kdv} and \emph{e.g.} adding counter-terms-like higher-derivative on $B$ may be a natural step. Finally, similarly to $2$d~\cite{Czech:2017ryf}, it would be interesting to give a clear interpretation (\emph{e.g.} counting gates) of different terms in higher-dimensional complexity action, as well as have a set of purely quantum computation arguments (\emph{e.g.} penalty factors for certain gates) for discarding some of the possible contributions. We will discuss and propose resolutions to some of these issues in what follows.
	
	\section{Uniformization and the Q-curvature Action} \label{sec:unif}
In this section we discuss a systematic and geometric way of interpreting the path integral optimization and the functional $I_{\Psi}[\phi,\hat{g}]$ in \emph{even-dimensional} CFTs using the Q-curvature action\footnote{The reason for restricting to even-dimensional CFTs is due to the fact that in odd-dimensions there is no trace-anomaly. However, one could potentially consider a square-root type (more generally, a DBI type) action \cite{Jatkar:2011ue, Bhattacharyya:2012tc} as a candidate for the (holographic) path integral complexity. We leave this as an interesting future avenue.}. We introduce the basic objects used in later discussions with a special focus on the Q-curvature, which is the higher-dimensional analogue of the Gauss curvature\footnote{For more details, we refer the readers to \cite{book}.}. We will see that higher-dimensional path integral complexity actions obtained from so-called uniformization problem, which we will also discuss, have a natural interpretation in terms of Q-curvature actions. Furthermore, we verify an essential property, namely the co-cycle condition, that must be satisfied in order for the Q-curvature action to be a valid path-integral complexity action. We also provide an intuitive explanation of the path integral optimization and connect it with the tensor network picture.

\subsection{Q-curvature}

	Consider a compact even-dimensional manifold $(\mathcal{M},\hat{g}_{ab})$ and a Weyl transformation of the metric: $\hat{g}_{ab} \rightarrow g_{ab} = e^{2 \phi(x)} \hat{g}_{ab}$, where $\phi(x)$ is a scalar function capturing the effect of the transformation. Under this transformation the Ricci scalar transforms as
	\begin{align}
		e^{2 \phi(x)} R (e^{2 \phi(x)} \hat{g}) = R(\hat{g}) - 2 (d-1) \Box_{\hat{g}} \phi(x) - (d-1) (d-2) |\nabla_{\hat{g}} \phi(x)|^2, \label{rictr}
	\end{align}
	where the subscript $\hat{g}$\footnote{Here by $\hat{g}$ or $g$, we indicate the metric itself, not the determinant of the metric.} indicates that the respective operators are evaluated on that metric, $d$ is the dimension of the manifold $\mathcal{M}$, and where $\Box_{\hat{g}}$ and $\nabla_{\hat{g}}$ are respectively the Laplace--Beltrami operator the covariant derivative with respect to $\hat{g}$. The notation $R(\hat{g})$ and $R(g)=R (e^{2 \phi(x)} \hat{g})$ means that the Ricci scalar has to be evaluated on the metrics $\hat{g}$ and $g = e^{2 \phi(x)} \hat{g}$ respectively. We also define a scalar $\mathcal{J}(g)$ by
	\begin{align}
	\mathcal{J}(g) = \frac{R(g)}{2 (d-1)}, \label{j}
	\end{align}
whose interpretation will be clear later on. The introduction of $\mathcal{J}(g)$ allows us to rewrite the transformation \eqref{rictr} as
\begin{align}
	e^{2 \phi(x)} \mathcal{J} (e^{2 \phi(x)} \hat{g}) = \mathcal{J}(\hat{g}) - \Box_{\hat{g}} \phi(x) - \left(\frac{d}{2} - 1 \right) |\nabla_{\hat{g}} \phi(x)|^2.
\end{align}
Specializing to $d=2$, the above transformation simplifies to
\begin{align}
	e^{2 \phi(x)} \mathcal{J} (e^{2 \phi(x)} \hat{g}) = \mathcal{J}(\hat{g}) - \Box_{\hat{g}} \phi(x). \label{go}
\end{align}
This relation is exactly equivalent to the \emph{Gauss-curvature prescription} \cite{book}
\begin{align}
	e^{2 \phi(x)} \mathcal{K} (e^{2 \phi(x)} \hat{g}) = \mathcal{K}(\hat{g}) - \Box_{\hat{g}} \phi(x), \label{gauss}
\end{align}
which shows how the Gauss curvature $\mathcal{K}(\hat{g})$ for the metric $\hat{g}$ changes under a Weyl transformation. Hence, in $d=2$, we identify $\mathcal{J}=\mathcal{K}$. One immediate question one can ask is whether there is an analogous generalized version of Eq.\eqref{go} in terms of higher-curvature invariants. To answer this, one defines the \emph{Schouten tensor}  for $d > 2$ as \cite{book}~\footnote{In $d$-dimensional manifolds ($d>2$) with locally conformally flat metrics (\emph{i.e.} with vanishing Weyl tensor) the curvature tensor is governed by the Schouten tensor $S_{ab}$.} 
\begin{align}
	S_{ab}(\hat{g}) = \frac{1}{d-2} \Big(R_{ab}(\hat{g}) - \mathcal{J} (\hat{g}) \, \hat{g}_{ab} \Big). \label{sch}
\end{align}
We are now in a position to define the Q-curvature. For a given metric $\hat{g}_{ab}$, the \emph{Branson Q-curvature} of order four in general dimensions $d > 4$ is defined as\footnote{In \cite{book}, the Q-curvature is defined with a negative sign. This is purely a matter of convention.}
\begin{align}
 \mathcal{Q}_{4,d}  (\hat{g})= -\frac{d}{2} \mathcal{J} (\hat{g})^2 + 2 \, S_{ab} (\hat{g}) S^{ab} (\hat{g}) + \Box_{\hat{g}}  \mathcal{J}(\hat{g}). \label{qdef}
\end{align}
Note that we have two indices in $\mathcal{Q}_{4,d}$. The first index denotes the order of the curvature and the second index represents the dimension. It is easy to see that $\Box_{\hat{g}}  \mathcal{J}(\hat{g})$ contains fourth-order derivatives of the given metric and hence $\mathcal{Q}_{4,d}$ also contains them. From now onwards, we often suppress the dependence of the metric for convenience. Using Eq.\eqref{j} and Eq.\eqref{sch}, we write the $\mathcal{Q}_{4,d}$ in a more convenient form
\begin{align}
	\mathcal{Q}_{4,d} = \frac{1}{2 (d-1)} \Box_{\hat{g}} R + \frac{2}{(d-2)^2} R_{ab} R^{ab} - \frac{d^2 (d-4) + 16 (d-1)}{8 (d-1)^2 (d-2)^2} R^2. \label{q4dim}
\end{align}
Our interest is $\mathcal{Q}_4 = \mathcal{Q}_{4,4}$, i.e., the $\mathcal{Q}_{4,d}$ in $4$-dimensions. Hence, from now onwards when we refer to the Q-curvature in $4$-dimensions, we mean $\mathcal{Q}_4 \equiv \mathcal{Q}_{4,4}$. Setting $d=4$ in Eq.\eqref{q4dim}, we obtain the expression of $\mathcal{Q}_4$ as
\begin{align}
	 \mathcal{Q}_4 =  \frac{1}{6} \Big(\Box_{\hat{g}} R + 3 R_{ab} R^{ab} - R^2 \Big).
\end{align}
Now, we come back to the question whether there is a generalization of Eq.~\eqref{go}. The answer is affirmative and we can directly generalise the Gauss-curvature prescription to the $\mathcal{Q}_4$-curvature prescription by the following theorem \cite{Branson1991ExplicitFD}.\\
\textbf{Theorem 1:} For a four-dimensional manifold equipped with a metric $\hat{g}_{ab}$, the $\mathcal{Q}_4$-curvature prescription states that the $\mathcal{Q}_{4}$ curvatures of conformally-related metrics satisfy
\begin{align}
	e^{4 \phi(x)} \mathcal{Q}_4 (e^{2 \phi(x)} \hat{g}) = \mathcal{Q}_4(\hat{g}) - \mathcal{P}_4 ({\hat{g}}) (\phi(x)), \label{y4}
\end{align}
where $\mathcal{P}_4 (\hat{g})$ is a differential operator given by
\begin{align}
	\mathcal{P}_4 (\hat{g}) = \Box_{\hat{g}}^2 + \nabla_a \big( 2 \mathcal{J} g^{ab} - 4 S^{ab} \big) \nabla_b. \label{p4}
\end{align}
Here $\mathcal{J}$ and $S_{ab}$ are defined by Eq.\eqref{j} and Eq.\eqref{sch} respectively. Note that this is the generalization of Eq.\eqref{go} or Eq.\eqref{gauss} to $4$-dimensions, where the Gauss curvature and the Laplace-Beltrami operator are replaced by the $\mathcal{Q}_4$ and $\mathcal{P}_4$ respectively. This result suggests that the Q-curvature is the generalization of Gauss curvature in higher dimensions.

Similar to $4$-dimensions, for a $2$-dimensional manifold equipped with a metric $\hat{g}_{ab}$, the $\mathcal{Q}_2$-curvature prescription states that
\begin{align}
	e^{2 \phi(x)} \mathcal{Q}_2 (e^{2 \phi(x)} \hat{g}) = \mathcal{Q}_2(\hat{g}) - \mathcal{P}_2 ({\hat{g}}) (\phi(x)).\label{y2}
\end{align}
This equation is nothing but Eq.\eqref{go} if one identifies $\mathcal{P}_2 (\hat{g}) \equiv \Box_{\hat{g}}$ and  $\mathcal{Q}_{2,d}$ as
\begin{align}
	\mathcal{Q}_{2,d} (\hat{g}) = \frac{R (\hat{g})}{2 (d-1)}  = \mathcal{J} (\hat{g}).
\end{align}
for $d\geq2$. In particular the second order Q-curvature in $2$-dimensions satisfies $\mathcal{Q}_2 \equiv \mathcal{Q}_{2,2} = R/2$, which immediately leads back to Eq.\eqref{rictr} in terms of $R(\hat{g})$. Hence $\mathcal{Q}_{2,d} \equiv \mathcal{J}$ for all $d$. 

With these operational definitions, we have encountered two differential operators namely $ \mathcal{P}_2$ and $ \mathcal{P}_4$, which are conformally covariant. The theorem below gives the transformation law of $\mathcal{P}_4$.\\
\textbf{Theorem 2:} Under the Weyl transformation $g_{ab} = e^{2 \phi(x)} \hat{g}$, the operator $\mathcal{P}_4(\hat{g})$ transforms according to
\begin{align}
	e^{4 \phi(x)}  \, \mathcal{P}_4( e^{2 \phi(x)} \hat{g}) = \mathcal{P}_4(\hat{g}), \label{p4c}
\end{align}
i.e., $\mathcal{P}_4 (\hat{g})$ is conformally covariant.\newline
\emph{Proof:} Using Theorem 1~\eqref{y4}, we write the LHS of the above equation as
\begin{align}
	e^{4 \phi}  \, \mathcal{P}_4( e^{2 \phi} \hat{g}) (\psi)  &= -e^{4 (\phi + \psi)}  \mathcal{Q}_4( e^{2 (\phi + \psi)} \hat{g}) + e^{4 \phi}  \mathcal{Q}_4( e^{2 \phi} \hat{g}), \nonumber \\
	&= \Big(-\mathcal{Q}_4(\hat{g}) + \mathcal{P}_4(\hat{g}) (\phi + \psi) \Big) - \Big(-\mathcal{Q}_4(\hat{g}) + \mathcal{P}_4(\hat{g}) (\phi) \Big), \nonumber \\
	&= \mathcal{P}_4(\hat{g}) (\psi),
\end{align}
which is the RHS of~\eqref{p4c}. Here the second line follows from~\eqref{y4} and in the third line, we have used the fact that $\mathcal{P}_4$ is a linear operator, and hence $\mathcal{P}_4(\hat{g}) (\phi + \psi) = \mathcal{P}_4(\hat{g}) (\phi) + \mathcal{P}_4(\hat{g}) (\psi)$, completing the proof.

By a similar argument, one can prove that $\mathcal{P}_2 (\hat{g})$ is also conformally covariant. i.e., under the conformal transformation $g = e^{2 \phi(x)} \hat{g}$, the operator $\mathcal{P}_2(\hat{g})$ transforms according to
\begin{align}
	e^{2 \phi(x)}  \, \mathcal{P}_2( e^{2 \phi(x)} \hat{g}) = \mathcal{P}_2(\hat{g}). \label{p2c}
\end{align}

In general, one could define a general form of the operator $\mathcal{P}_{2,d}$ in $d$-dimensions which is known as the \emph{Yamabe operator}
\begin{align}
	\mathcal{P}_{2,d} (\hat{g}) = \Box_{\hat{g}} - \bigg(\frac{d}{2} - 1 \bigg) \mathcal{Q}_{2,d} (\hat{g}) = \Box_{\hat{g}} - \frac{d-2}{4 (d-1)}  R (\hat{g}).
\end{align}
Similar to the Q-curvature, here the first index denotes the order of the curvature while the second one indicates the dimension. One can define $\mathcal{P}_{4,d}$ for $d \geq 2$, known as the \emph{Paneitz operator} \cite{Riegert:1984kt, Eastwood1985ACI, 2008}. It is defined as
\begin{align}
	\mathcal{P}_{4,d} (\hat{g}) = \Box_g^2 + \nabla_a \big( (d-2) J g^{ab} - 4 S^{ab} \big) \nabla_b - \bigg(\frac{d}{2} - 2 \bigg) \mathcal{Q}_{4,d}(\hat{g}) .
\end{align}
Note that for $d=4$ this reduces to Eq.\eqref{p4}.

The important aspect of the Yamabe and Paneitz operators is that they are conformally covariant \cite{book} i.e.,
\begin{align}
		e^{(\frac{d}{2} +1)\phi}  \, \mathcal{P}_{2,d} ( e^{2 \phi} \hat{g}) (\psi) &= \mathcal{P}_{2,d}(\hat{g}) (e^{(\frac{d}{2} -1)\phi} \psi ), \label{pan1}\\
		e^{(\frac{d}{2} +2)\phi}  \, \mathcal{P}_{4,d} ( e^{2 \phi} \hat{g}) (\psi) &= \mathcal{P}_{4,d}(\hat{g}) (e^{(\frac{d}{2} -2)\phi} \psi ), \label{pan2}
\end{align}
Note that the above covariance property reduces to Eq.\eqref{p2c} and Eq.\eqref{p4c} for $d=2$ and $d=4$ respectively.

The generalization of Theorem 1~\eqref{y4} to general dimensions is straightforward (see \emph{e.g.}~\cite{book}). For an even $d$-dimensional manifold equipped with a metric $\hat{g}_{ab}$ the following identity holds
\begin{align}
	e^{d \phi(x)} \mathcal{Q}_d (e^{2 \phi(x)} \hat{g}) = \mathcal{Q}_d (\hat{g}) - \mathcal{P}_d ({\hat{g}}) (\phi(x)), \label{yd}
\end{align}
where $\mathcal{Q}_{d}\equiv \mathcal{Q}_{d,d}$ and $\mathcal{P}_{d}\equiv\mathcal{P}_{d,d}$ are higher-dimensional generalizations of the $\mathcal{Q}_{2,d}$ and $\mathcal{Q}_{4,d}$ Q-curvatures and the Yamabe $\mathcal{P}_{2,d}$ and Paneitz $\mathcal{P}_{4,d}$ operators and are respectively known as Branson's Q-curvature and the Graham--Jenne--Mason--Sparling (GJMS) operator. We will return to these objects in the following section. A proof of~\eqref{yd} for even-dimensional Riemannian manifolds, known as the fundamental identity theorem, is given in \cite{book}. This identity leads to the following theorem.\\
\textbf{Theorem 3:} For an even $d$-dimensional manifold, the following functional
\begin{align}
	\mathcal{T}_d (g) = \int_{\mathcal{M}^d} \mathcal{Q}_d(\hat{g})\, \mathrm{vol}(\hat{g}),~\label{thrm3}
\end{align}
is  invariant under conformal transformations.\\
\emph{Proof:} First, we write
\begin{align}
	\mathcal{T}_d (e^{2 \phi} \hat{g}) = \int_{\mathcal{M}^d} \mathcal{Q}_d(e^{2 \phi} \hat{g})\, \mathrm{vol}(e^{2 \phi} \hat{g}) = \int_{\mathcal{M}^d} e^{d \phi} \, \mathcal{Q}_d(e^{2 \phi} \hat{g})\, \mathrm{vol}( \hat{g}), 
\end{align}
where $\mathrm{vol}(\hat{g})$ is a convenient notation for $\sqrt{\hat{g}}$  and where we have neglected the overall multiplicative factor. The second equality follows from the fact that, under rescaling, $\mathrm{vol}(\hat{e^{2 \phi}\hat{g}}) = e^{d \phi} \,\mathrm{vol}(\hat{g})$ for  $d$-dimensions. Now, using Eq.\eqref{yd}, we can write this as
\begin{align}
	\mathcal{T}_d (e^{2 \phi} \hat{g})  = \int_{\mathcal{M}^d} \big[\mathcal{Q}_d (\hat{g}) - \mathcal{P}_d ({\hat{g}}) (\phi(x))\big] \mathrm{vol}( \hat{g}).
\end{align}
It has been shown in \cite{book} that the integral over $\mathcal{P}_d$ vanishes. This implies
\begin{align}
	\mathcal{T}_d (e^{2 \phi} \hat{g})  = \int_{\mathcal{M}^d} \mathcal{Q}_d (\hat{g}) \,  \mathrm{vol}( \hat{g}) = \mathcal{T}_d (\hat{g}) 
\end{align}
completing the proof. 
Equipped with these definitions, we now state the \emph{Yamabe problem}~\cite{Lee1987TheYP}.\\
\newline
\textbf{Yamabe problem (in $2d$ and $4d$):} Consider a $2$- and $4$-dimensional manifold equipped with a metric $\hat{g}_{ab}$, and a Weyl transformation $g = e^{2 \phi(x)} \hat{g}$, which defines an equivalence class of conformally-equivalent metrics $[g]$. Can we find a class of metrics which have a constant Q-curvature $\mathcal{Q}_2$ and $\mathcal{Q}_4$ in $d=2$ and $d=4$ respectively?

To state this problem more clearly, consider Eqs.\eqref{y2} and \eqref{y4}. The Yamabe problem demands that the Q-curvatures of the Weyl-rescaled metric should be constant, i.e., $\mathcal{Q}_2 (e^{2 \phi} \hat{g}) = \Lambda_2$ and $\mathcal{Q}_4 (e^{2 \phi} \hat{g}) = \Lambda_4$. Here, we look for constants $\Lambda_2, \Lambda_4$ which are negative \emph{i.e.}, we want to find a class of conformal transformation for which $\Lambda_2, \Lambda_4 < 0$. In these cases, Eqs.\eqref{y2} and \eqref{y4} are simplified to
\begin{align}
	 &\mathcal{Q}_2(\hat{g}) - \mathcal{P}_2 ({\hat{g}}) (\phi)=\Lambda_{2}\,e^{2 \phi}~, \label{yc2} \\
	 &\mathcal{Q}_4(\hat{g}) - \mathcal{P}_4 ({\hat{g}}) (\phi)=\Lambda_{4}\,e^{4 \phi}~. \label{yc4}
\end{align}
The above equations can be recast as a variational problem, \emph{i.e.} one can view them as the Euler-Lagrange equations obtained by the variation of the action
	\begin{align} 
	I_{d}[\phi,\hat{g}] = k  \int_{\mathcal{M}^{d}} \mathrm{d}^d x  \Big(\phi \,\mathcal{P}_{d}(\hat{g}) \, \phi - 2 \mathcal{Q}_{d} (\hat{g}) \, \phi + \frac{2}{d} \Lambda_d \,e^{d \phi} \Big) \,  \mathrm{vol}( \hat{g}), \label{comd0}
\end{align}
where $\mathcal{Q}_{d} \in \{\mathcal{Q}_2, \mathcal{Q}_4 \}$ and $\mathcal{P}_{d} \in \{\mathcal{P}_2, \mathcal{P}_4 \}$ are the Q-curvature and Yamabe/Paneitz operators in $d=2$ and $d=4$ respectively,\footnote{The negative sign before $\mathcal{Q}_d$ is arbitrary. It depends on how we choose the definition of Q-curvature, for example in \eqref{qdef}.} $k$ is a proportionality constant, and $\Lambda_d \in \{\Lambda_2, \Lambda_4 \}$ are the (negative) constants Q-curvature of the Weyl-rescaled metric.

Note that~\eqref{yc2} can be re-written in terms of the Weyl-rescaled metric $e^{2\phi}\hat{g}_{ab}$ as
\begin{align}
		R(e^{2\phi}\hat{g})=2\mathcal{Q}_{2}(e^{2\phi}\hat{g})=2\Lambda_{2}<0~,\label{eq:LiouR}
\end{align}
which is nothing else than the Liouville equation discussed previously and cast as in terms of the Ricci scalar as in~\eqref{eq:LiouvilleEq}. This implies that~\eqref{yc4}, written in terms of the Weyl rescaled metric as $\mathcal{Q}_4 (e^{2 \phi} \hat{g}) = \Lambda_4$, can be regarded as a natural generalization of the Liouville equation in $d=4$.

It is illustrative to find interesting solutions of Eq.\eqref{yc2} and \eqref{yc4}. For convenience, we choose the reference metric $\hat{g}$ as Euclidean flat $\hat{g}_{ab}=\delta_{ab}$. This implies that $\mathcal{Q}_2$ and $\mathcal{Q}_4$ vanish identically, and $\mathcal{P}_2 = \Box = \partial^2$ and $\mathcal{P}_4 = \Box^2 = \partial^4$. Hence in this case the equations~\eqref{yc2}~\eqref{yc4} simplify to
\begin{align}
	&\partial^2 \phi=\Theta_{2}\,e^{2 \phi}\,, \label{ycs2} \\
	&\partial^4 \phi= \Theta_{4}e^{4 \phi} \,, \label{ycs4}
\end{align}
where $\Theta_2= -\Lambda_2$ and $\Theta_4 = -\Lambda_4$ respectively. Along with the boundary condition $e^{2 \phi(-\tau = \epsilon,x)} = 1/\epsilon^2$, where ($-\tau$) is the Euclidean time, the solution to both equations is given by
\begin{align}
	e^{2 \phi(\tau,x)} =  \frac{1}{\tau^2}~\label{eq:ycshyp}
\end{align}
for $d=2, 4$ with $\{\Theta_2, \Theta_4\} = \{1, 6\}$ respectively. We will again come across this fact later on. The above discussion directly leads to the \emph{uniformization problem} of conformally-equivalent metrics which we will discuss in the following section.

\subsection{Path Integral Optimization as a Uniformization Problem}
	
	In conformal geometry, one can formulate the following \emph{uniformization} problem~\cite{Cercle:2019jxx}: given a reference metric $\hat{g}_{ab}$, can one find a metric $g_{ab}$ with a constant (negative) Q-curvature $\Lambda_d < 0$ that is conformally equivalent to $\hat{g}_{ab}$? The answer turns out to be affirmative and the required metric can be found by extremizing the \emph{Q-curvature action} in $d$ even dimensions, which is given by~\cite{Levy:2018bdc, Cercle:2019jxx}
	\begin{align} 
		I_{d}[\phi,\hat{g}] = \frac{d}{2 \Omega_d (d-1)!} \int_{\mathcal{M}^d} \mathrm{d}^d x  \Big(\phi \,\mathcal{P}_{d} (\hat{g}) \, \phi - 2 \mathcal{Q}_{d}(\hat{g}) \, \phi + \frac{2}{d} \Lambda_d \,e^{d \phi} \Big) \,  \mathrm{vol}( \hat{g}), \label{comd}
	\end{align}
	where $\Omega_d = 2 \pi^{(d+1)/2}/\Gamma[(d+1)/2]$ is the $d$-dimensional volume of sphere $S^{d}$, the conformally covariant differential operator $\mathcal{P}_{d}$ is the aforementioned GJMS operator~\cite{Graham:1992jms} and $\mathcal{Q}_{d}$ is the Q-curvature scalar defined for the reference metric $\hat{g}_{ab}$.\footnote{Note that $\mathcal{Q}_{\hat{g}}$ is the Q-curvature of the reference metric whereas $\Lambda$ is the Q-curvature of the ``uniformised'' metric.} These objects are generalizations of the Laplace-Beltrami operator and Gauss curvature respectively from $2$-dimensions to higher even-dimensions and transform in a similar way as their $2$-dimensional counterparts, as we have seen in previous sections. In particular, $\mathcal{P}_d ({\hat{g}})$ is the generalization of the Yamabe and Paneitz operators in even $d$-dimensions. The form of their leading structure is given by
	\begin{align} \label{eq:defQP}
	\mathcal{P}_{d} = \Box^{d/2} + \mathrm{lower ~order}, ~~~
	\mathcal{Q}_{d} = \frac{1}{2(d-1)} \Box^{\frac{d}{2}-1} R + \cdots
	\end{align}
	where $d$ is any even dimension and where the ellipsis denotes higher-curvature invariants constructed from the Ricci scalar and tensor as well as from their derivatives. Their explicit forms in $2$- and $4$-dimensions are given in the previous section.
		
	For a hyperbolic metric~\eqref{eq:EucAdSPoinc} in $d=2$ and $d=4$, one obtains $\mathcal{Q}_2 = -1$ and $\mathcal{Q}_4 = -6$ respectively, which we will use later (they will be denoted by $\Lambda$ since they are the Q-curvatures of the optimized metric). 
	The transformation properties of the GJMS operator $\mathcal{P}_{d}$ allows us to construct conformal invariant quantities, which we have encountered in the previous section. For example, for an even-dimensional conformally flat manifold $\mathcal{M}$, the integral of the Q-curvature yields
	\begin{align} \label{eq:Euler}
	\mathcal{T}_{d}(\hat{g})=\int_{\mathcal{M}^{d}}   \mathcal{Q}_{d}(\hat{g}) \,  \mathrm{vol}( \hat{g}) = \frac{1}{2} \Omega_d (d-1)! \, \chi(\mathcal{M}), 
	\end{align}
	where $\Omega_d = 2 \pi^{(d+1)/2}/\Gamma[(d+1)/2]$ is the $d$-dimensional volume of sphere $S^{d}$ and $\chi(\mathcal{M})$ is known as the \emph{Euler characteristic} of $\mathcal{M}$. It is easy to see that in $d=4$, the invariant is $8 \pi^2 \chi(\mathcal{M})$. In general, Eq.\eqref{eq:Euler} will be supplemented by an integral over the Weyl tensor \cite{book}, which is conformally invariant for any dimensions $d > 2$, as we showed in~\eqref{thrm3}. We will be particularly interested in conformally-flat spacetimes, in which case this expression vanishes identically.
	
	 Moreover, one can regard~\eqref{comd} as the higher even-dimensional version of the Liouville action~\eqref{eq:LiouvilleAct}. The equation of motion obtained from the variation of the Q-curvature action~\eqref{comd} is given by
	\begin{align}\label{OptHigherDimConstr}
		\mathcal{P}_d (\hat{g}) \,\phi - \mathcal{Q}_d (\hat{g})=  \Theta_d\, e^{d \phi},
	\end{align}
	where $\Theta_d = -\Lambda_d > 0$ is the cosmological constant. This is the higher-dimensional version of Liouville equation~\eqref{eq:LiouvilleEq} for even-dimensional manifolds.

	For convenience and along the lines of path integral optimization we take our reference metric to be the Euclidean flat metric, as we did at the end of the last section. This gives $\mathcal{Q}_d = 0$ and the GJMS operator reduces to $\mathcal{P}_{d} = \Box^{d/2} = \partial^d$, resulting in the equation of motion
	\begin{align}
		\partial^d \phi = \Theta_d\, e^{d \phi}. \label{dimd}
	\end{align}
	Along with the  boundary condition 
	\begin{align}\label{eq:BdryCond}
	e^{2 \phi(-\tau = \epsilon,x)} = \frac{1}{\epsilon^2}~,
	\end{align}
	where ($-\tau$) is the Euclidean time, the solution is given by
	\begin{align}
		e^{2 \phi(\tau,x)} = \bigg[\frac{(d-1)!}{\Theta_d} \bigg]^{2/d} \frac{1}{\tau^2}, \label{musol0}
	\end{align}
	confirming the optimal geometry as hyperbolic \footnote{This hyperbolic solution was shown to be a minimum of the Liouville action in \cite{Caputa:2017yrh}. In higher dimensions we have not been able to prove that this is the lower bound of the Q-curvature action. Still, the Q-curvature action may still be a meaningful measure of complexity since the optimization procedure should be stoped when $e^\phi\sim O(1)$  such that we cannot coarse-grain more than the original lattice. We thank Tadashi Takayanagi for comments on this issue. }. This solution can be rewritten in terms of a parameter $\mu = [\Theta_d\,/(d-1)!]^{2/d}$ simply as
	\begin{align}
		e^{2 \phi(\tau,x)} =  \frac{1}{\mu \tau^2}~. \label{musol1}
	\end{align}

	This result can be directly linked to the discussion of path integral optimization and the optimization of the hyperbolic metrics~\eqref{eq:PoincDiscH2}. As the optimized metric corresponds to $\Theta_d= (d-1)!$, the optimization condition is given by $\mu =1$. For example, in $2$-dimensions we readily obtain $\Theta_2 = \mu =1$ for the optimized metric and the optimized geometry is the Poincar\'{e} half-plane. Alternatively, this implies $\Lambda_2 = -1$, corresponding to the Gaussian curvature of the Poincar\'{e} half-plane.
		In $4$-dimensions, the optimized geometry $\mu =1$ corresponds to $\Theta_4 = 6$. This again implies $\Lambda_2 = -6$ which is the Q-curvature of the optimized hyperbolic geometry in $4$-dimensions. 
	
	Note that the value of the cosmological constant $\Theta_d$ is \emph{automatically} fixed according to the spacetime dimension and physically corresponds to the (negative) Q-curvature of the optimal geometry. This also explains why we need to set $\mu =1$ for the optimized geometry. This result intuitively suggests that the amount of Q-curvature of the the optimal (hyperbolic) geometry sets the scale of the optimization and the boundary geometry automatically picks up the optimal way of performing the path integration. It is natural to follow the analogy and propose that the corresponding path integral complexity is then given by the on-shell value of the Q-curvature action, which for $d$-dimensions behaves as $\sim V_{d-1}/\epsilon^{d-1}$, where $V_{d-1}$ is the $(d-1)$-dimensional spatial volume, consistent with the holographic ``complexity=volume'' proposal.

	It is also instructive to verify the optimization constraint~\cite{Boruch:2021hqs} in the context of the Q-curvature. In $2$-dimensions, the optimization constrain reads $R^{(2)} =  -2\mu$. From Eq.\eqref{j} (note that $\mathcal{J}=\mathcal{Q}_2$), we obtain $\mathcal{Q}_2 = R^{(2)}/2$, which implies that the optimization constraint $\mu =1$  corresponds to the optimized Q-curvature $\mathcal{Q}_2 =-1$, which is the result for the hyperbolic geometry. In higher dimensions the optimization constraint is instead given by~\eqref{OptHigherDimConstr}. In $4$-dimensions, this optimization corresponds to $R^{(4)} = -12 \mu$, and $\mathcal{Q}_4 = -6$, which is the optimized Q-curvature for the geometry in $4$-dimensions, that was discussed previously. This holds for all even-dimensions. As a consequence, the optimization constraint is naturally incorporated within the uniformization formulation via the Q-curvature action.
	
	\subsection{Improved Q-curvature Action and the Co-cycle Condition}
	The Liouville action has a number of interesting properties. For example, an improved version of the Liouville action has been defined in \cite{Caputa:2017yrh} by subtracting the potential term proportional to the volume in \eqref{eq:HigherDPathIntAction}, which satisfies the so-called \emph{co-cycle condition} \cite{Boruch:2021hqs}
	\begin{align}
		I_{\Psi}[g_1, g_2] + 	I_{\Psi}[g_2, g_3] = 	I_{\Psi}[g_1, g_3], \label{cocycle}
	\end{align}
	where $I_{\Psi}[g_1, g_2]$ computes the complexity between two TNs described by metrics $g_1$ and $g_2$. It has been argued that a legitimate path integral complexity action (in any dimension) should obey this co-cycle condition. Hence, it is important to verify whether the Q-curvature action defined in Eq.\eqref{comd} also satisfies this condition.\\
	\newline
	\textbf{Claim:} The improved Q-curvature action
	\begin{align}
		I_{d}^{\mathrm{im}}[\phi,\hat{g}] = I_{d} [\phi,\hat{g}] -  I_{d} [0,\hat{g}].
	\end{align}  \label{eq:ImprovedQaction1}
 obeys the cocycle relation \eqref{cocycle} where $ I_{d} [\phi,\hat{g}]$ is given by Eq.\eqref{comd}.
	 
\emph{Proof:} First, we separate $I^{\mathrm{im}}_d[\phi,\hat{g}]$ into two parts 
	\begin{align}
		I_{d}^{K}[\phi,\hat{g}] &=  \int_{\mathcal{M}^d}  \big(\phi \mathcal{P}_{d} (\hat{g}) \, \phi - 2 \mathcal{Q}_{d}(\hat{g}) \, \phi  \big) \, \mathrm{vol}(\hat{g}),  \label{first2} \\
			I_d^V [e^{2 \phi} \hat{g}, \hat{g}] &= \int_{\mathcal{M}^d} (e^{d \phi} - 1) \,  \mathrm{vol} (\hat{g}),\label{pot00}
	\end{align}
	and ignore the overall constant $d/2 \Omega_d (d-1)!$ that will not play any role in this proof. We then separately show that each of the above terms satisfy the co-cycle condition. For convenience, we show the proof for the first term \eqref{first2} in $d=4$, but it can be generalized to any even dimensions. We closely follow the method outlined in \cite{book}. 
	
Let us then consider the following action
	\begin{align}
			I_{4}^{K}[\phi,\hat{g}] = \int_{\mathcal{M}^4}  \big(\phi \mathcal{P}_{4} (\hat{g}) \, \phi - 2 \mathcal{Q}_{4}(\hat{g}) \, \phi  \big) \, \mathrm{vol}(\hat{g}). \label{im}
	\end{align}
From Theorem 1~\eqref{y4}, we obtain
		\begin{align}
			e^{4 \phi} \mathcal{Q}_4 (e^{2 \phi} \hat{g}) = \mathcal{Q}_4(\hat{g}) - \mathcal{P}_4 ({\hat{g}}) (\phi), \label{co1}
		\end{align}
	where we suppress the coordinate dependence of $\phi$. Adding a term $\mathcal{Q}_4 (\hat{g})$ to both sides and multiplying them by $\phi$ and $\mathrm{vol}(\hat{g})$, the above equation can be written as
	\begin{align}
		\phi \big[2\mathcal{Q}_4(\hat{g}) - \mathcal{P}_4 ({\hat{g}}) (\phi) \big] \mathrm{vol}(\hat{g}) = \phi \big[ \mathcal{Q}_4(\hat{g}) \, \mathrm{vol}(\hat{g}) + \mathcal{Q}_4(e^{2 \phi}\hat{g}) \, \mathrm{vol}(e^{2 \phi}\hat{g}) \big], \label{co2}
	\end{align}
	where we have used the fact that $\mathrm{vol}(\hat{e^{2 \phi}\hat{g}}) = e^{4 \phi} \,\mathrm{vol}(\hat{g})$. Hence Eq.\eqref{im} can be re-written as the integral
	\begin{align}
		\mathcal{S}[e^{2 \phi}\hat{g}, \hat{g}] = \int_{\mathcal{M}^4} \phi \big[ \mathcal{Q}_4(\hat{g}) \, \mathrm{vol}(\hat{g}) + \mathcal{Q}_4(e^{2 \phi}\hat{g}) \,  \mathrm{vol}(e^{2 \phi}\hat{g}) \big], \label{sint}
	\end{align}
and we define its integrand as
 	\begin{align}
 	\mathcal{L}[e^{2 \phi}\hat{g}, \hat{g}] =  \phi \big[ \mathcal{Q}_4(\hat{g}) \,  \mathrm{vol}(\hat{g}) + \mathcal{Q}_4(e^{2 \phi}\hat{g})  \, \mathrm{vol}(e^{2 \phi}\hat{g}) \big].
 \end{align}
From the transformation rules of its elements, it can be shown that the integrand satisfies the following identity
	\begin{align}
	\mathcal{L}[e^{2 (\phi + \psi)}\hat{g}, e^{2 \phi} \hat{g}] &=  \psi \big[ \mathcal{Q}_4(e^{2 \phi} \hat{g}) \, \mathrm{vol}(e^{2 \phi} \hat{g}) + \mathcal{Q}_4(e^{2 (\phi + \psi)}\hat{g}) \, \mathrm{vol}(e^{2 (\phi + \psi)}\hat{g}) \big]. \nonumber \\
	&= 	\psi \big[2\mathcal{Q}_4(e^{2 \phi} \hat{g}) - \mathcal{P}_4 (e^{2 \phi} {\hat{g}}) (\psi) \big] \mathrm{vol}(e^{2 \phi} \hat{g}) ,
\end{align}
where in the second line we have used Eq.\eqref{co2}. Now using $\mathrm{vol}(\hat{e^{2 \phi}\hat{g}}) = e^{4 \phi} \,\mathrm{vol}(\hat{g})$, we write
	\begin{align}
	\mathcal{L}[e^{2 (\phi + \psi)}\hat{g}, e^{2 \phi} \hat{g}] &= 	\psi \big[2 e^{4 \phi} \mathcal{Q}_4(e^{2 \phi} \hat{g}) - e^{4 \phi} \mathcal{P}_4 (e^{2 \phi} {\hat{g}}) (\psi) \big] \mathrm{vol}( \hat{g}) \nonumber \\
	&= \psi \big[2\mathcal{Q}_4( \hat{g}) - 2 \mathcal{P}_4 ( {\hat{g}}) (\phi) - \mathcal{P}_4 ( {\hat{g}}) (\psi) \big] \mathrm{vol}( \hat{g}), \label{first}
\end{align}
where in the last line we have used Eq. \eqref{y4} and Eq.\eqref{p4c}. Similarly one can write
	\begin{align}
	\mathcal{L}[e^{2 \phi}\hat{g}, \hat{g}] = \phi \big[2\mathcal{Q}_4( \hat{g}) - \mathcal{P}_4 ( {\hat{g}}) (\phi) \big] \mathrm{vol}( \hat{g}), \label{sec}
\end{align}
Hence, by adding Eq.\eqref{first} and Eq.\eqref{sec}, we obtain
\begin{align}
\mathcal{L}[e^{2 (\phi + \psi)}\hat{g}, e^{2 \phi} \hat{g}] + \mathcal{L}[e^{2 \phi}\hat{g}, \hat{g}] &= (\phi + \psi) \big[2\mathcal{Q}_4( \hat{g}) - \mathcal{P}_4 ( {\hat{g}}) (\phi + \psi) \big] \mathrm{vol}( \hat{g}) \nonumber \\
&+ [\phi \mathcal{P}_4(\hat{g}) \psi - \psi \mathcal{P}_4(\hat{g}) \phi ] \mathrm{vol}( \hat{g}).
\end{align}
The first term is $\mathcal{L}[e^{2 (\phi + \psi)}\hat{g}, \hat{g}]$, and the second term can be  written as a total derivative term, which can be neglected \cite{book}\footnote{In fact, by analogy with boundary Liouville action, one should be able to derive (or do a deeper search of the math literature) the boundary Q-curvature and we leave this analysis for as an interesting future problem.}. Hence, the integral \eqref{sint} yields,
\begin{align}
	\mathcal{S}[e^{2 (\phi + \psi)}\hat{g}, e^{2 \phi} \hat{g}] + \mathcal{S}[e^{2 \phi}\hat{g}, \hat{g}] &= \mathcal{S}[e^{2 (\phi + \psi)}\hat{g}, \hat{g}], 
\end{align}
which further implies
\begin{align}
I_4^{K}[e^{2 (\phi + \psi)}\hat{g}, e^{2 \phi} \hat{g}] + I_4^{K}[e^{2 \phi}\hat{g}, \hat{g}] &= I_4^{K}[e^{2 (\phi + \psi)}\hat{g}, \hat{g}].
\end{align}
Choosing $g_1 = e^{2 (\phi + \psi)}\hat{g}, ~g_2 =  e^{2 \phi} \hat{g}$ and $g_3 = \hat{g}$, this proves our claim. 

The proof for the potential term can be done in general even dimensions. From the definition \eqref{pot00}, we readily verify the identity
\begin{align}
	I_d^V[e^{2 (\phi + \psi)}\hat{g}, e^{2 \phi} \hat{g}] + I_d^V [e^{2 \phi}\hat{g}, \hat{g}] &= I_d^V [e^{2 (\phi + \psi)}\hat{g}, \hat{g}].
\end{align}
i.e., the action $I_d^V$ obeys the co-cycle condition. Hence, as we argued before, the improved action defined by
\begin{align}
	I_{d}^{\mathrm{im}}[\phi,\hat{g}] &= \frac{d}{2 \Omega_d (d-1)!} \, (I_d^K + I_d^V) \nonumber \\
	&=  \frac{d}{2 \Omega_d (d-1)!}  \int_{\mathcal{M}^d}  \big(\phi \mathcal{P}_{d} (\hat{g}) \, \phi - 2 \mathcal{Q}_{d}(\hat{g}) \, \phi + (e^{d \phi} - 1)  \big) \, \mathrm{vol}(\hat{g}),
\end{align}
will also satisfy the co-cycle condition. Therefore the full improved Q-curvature action $I_{d}^{\mathrm{im}}[\phi,\hat{g}] $ obeys the co-cycle condition and is a legitimate candidate for a path-integral complexity action in even-dimensional spacetimes.
	
\subsection{Q-curvature vs Higher-Dimensional Complexity Action}\label{sec:asym}

Let us now discuss the difference between the optimization with the Q-curvature action and the action \eqref{eq:HigherDPathIntAction}. As we have described in the previous sections, in $d=2$ dimensions the Q-curvature $\mathcal{Q}_{2}$ and the Ricci scalar $R^{(2)}$ are directly related to each other, and hence provide the same amount of information about the curvature of the $2$-dimensional geometry. This is not surprising since $2$-dimensional orientable manifolds can be characterized by a single function corresponding to the conformal (Weyl) factor leading to conformally-equivalent metrics characterized by the Gauss curvature $\mathcal{K}$. This statement is not true for higher dimensions and in general the Ricci scalar $R$ and Q-curvature of an even-dimensional orientable manifold will contain different information about its curvature.

	It is relevant to note that if a geometry has a constant Ricci scalar curvature $R=$ const, then it will \emph{not} necessarily have a constant Q-curvature given that the latter also depends generically on other curvature invariants, as can be seen by Eq.\eqref{eq:defQP}. In this sense, $R=$ const is a \emph{weaker} condition on the curvature of a manifold than $\mathcal{Q}_{d}=$ const. At the same time, one can view the condition $\mathcal{Q}_{d}=$ const as leading to a $d$-th order differential equation on the metric for an even $d$-dimensional manifold (see Eq.\eqref{dimd}), whereas $R=$const will \emph{always} lead to a second-order differential equation regardless of the dimension of the manifold. This also implies that solving $\mathcal{Q}_{d}=$ const in higher number of dimensions rapidly becomes a challenging task.
	
	The first case where the difference between the Ricci scalar and the Q-curvature can be manifestly seen is in $4$-dimensions. For simplicity, consider an conformally flat $4$-dimensional manifold whose metric $g_{ab}$ written in local coordinates $x\in\lbrace \tau,\vec{x}\rbrace$ is given by
	\begin{align} \label{eq:CFlat4d}
		\textrm{d}s^{2}=e^{2\phi(x)}\left(\textrm{d}\tau+\textrm{d}\vec{x}\right)~.
	\end{align}
	Assuming that the conformal factor $\phi(x)$ depends only on one spacetime coordinate, say $\tau$, we find that the Q-curvature and Ricci scalar $R$ of the conformally flat metric~\eqref{eq:CFlat4d} are respectively given by
	\begin{align}
		\mathcal{Q}_{4}=-e^{-4\phi(\tau)} \partial_{\tau}^{4}{\phi}(\tau)~,~~~~~R^{(4)}=-6e^{-2\phi(\tau)}\left(\partial_{\tau}^{2}{\phi}(\tau)+\left(\partial_{\tau}\phi(\tau)\right)^{2}\right)~,
	\end{align}	
	A solution of the $\mathcal{Q}_{4}=-\Theta\,=$ const.$<0$ equation with boundary condition~\eqref{eq:BdryCond} is given by
	\begin{align} \label{eq:Q4const}
		\phi(\tau)=\log\left(\frac{6^{1/4}}{\Theta^{1/4}\,\,(\tau+\epsilon)+6^{1/4}\epsilon}\right)~.
	\end{align}
	This solution corresponds to a constant Ricci scalar given by $R^{(4)}=-2\sqrt{6\Theta\,}$, which is consistent with the observation in the previous section. This hyperbolic metric would hence be the solution to the uniformization problem and therefore also to the path integral optimization problem cast in therms of the Q-curvature, consistent with the expectations from AdS/TN considerations.
	
	If we instead focus on the constraint $R^{(4)}=-12\mu=$ const., with $0<\mu\leq 1$, we can look for a solution of the differential equation
	\begin{align}\label{eq:Ricciconstd4}
		2\,\mu\,e^{2\phi(\tau)}=(\partial_{\tau}\phi(\tau))^2+\partial_{\tau}^{2}\phi(\tau)~,
	\end{align}
	of the form
	\begin{align}\label{eq:Ricciconstd4Ansatz}
		\phi(\tau)=\log\left(\frac{1}{\xi(\tau)}\right)~,
	\end{align}
	with $\xi(\tau)>0$ a real and smooth function satisfying the boundary condition $\xi(-\epsilon)^{2}=\epsilon^{2}$. Indeed such a solution can be found and is given by
	\begin{align}\label{eq:Ricciconstd4Sol}
		\xi(\tau)= (-1)^{1/4}\,e^{-\kappa/2}\,\mu^{1/4}\,\textrm{sn}\left((-1)^{3/4}\,e^{\kappa/2}\,\mu^{1/4}\,\tau+\tau_{0}\,,\,-1\right)~,
	\end{align}
	where
	\begin{align}
		\tau_{0}=(-1)^{3/4}e^{\kappa/2}\mu^{1/4}\epsilon+\textrm{arcsn}\left(\frac{(-1)^{3/4}e^{\kappa/2}\epsilon}{\mu^{1/4}}\,,\,-1\right)~,
	\end{align}
	and where sn$(u\,, \,m)$ is a Jacobi elliptic function, arcsn$(v\,,\,m)$ is its inverse function and where $\kappa$ is a constant of integration that can be determined by imposing a condition on $\partial_{\tau}\phi(\tau)\vert_{\tau=-\epsilon}$. 
	The Jacobi elliptic function sn$(u\,, \,m)$ is a meromorphic function in both arguments which is doubly periodic in $u$ with periods $4K(m)$ and $2$\,i\,$K(1-m)$, where $K$ is the \emph{complete} elliptic integral of the first kind given by $K(m):=F(\pi/2\vert m)$, where $F(\varphi\vert m):=\int^{\varphi}_{0}\textrm{d}\,\theta\,\left(1-m\sin(\theta)^{2}\right)^{-1/2}$ is the (incomplete) elliptic integral of the first kind. To be precise, sn$(u\,, \,m):=\sin(\textrm{am}(u\,\vert\,m))$, where am$(u\,\vert\,m)$ is the \emph{amplitude} of the Jacobi elliptic functions, defined as the inverse of the incomplete elliptic integral of the first kind: am$(u\,\vert\,m)\equiv F^{-1}(u\,\vert\,m)$.
	Solution~\eqref{eq:Ricciconstd4Sol} leads to a Weyl factor of the form
	\begin{align}\label{eq:phiRicciconst}
		\phi(\tau)=\log\left[\frac{e^{\kappa/2}}{(-1)^{1/4}\,\mu^{1/4}\,\textrm{sn}\left((-1)^{3/4}\,e^{\kappa/2}\,\mu^{1/4}\,\tau+\tau_{0}\,,\,-1\right)}\right]~,
	\end{align}
	As a consequence, this solution to the $R^{(4)}=-12\mu=$ const equation leads to a metric of the form
	\begin{align}\label{eq:RconstMetric}
		\textrm{d}s^{2}=\frac{e^{\kappa}\left(\textrm{d}\tau^{2}+\textrm{d}\vec{x}\right)}{(-1)^{1/2}\,\mu^{1/2}\,\textrm{sn}^{2}\left((-1)^{3/4}\,e^{\kappa/2}\,\mu^{1/4}\,\tau+\tau_{0}\,,\,-1\right)}~,
	\end{align}
	which indeed satisfies the boundary condition~\eqref{eq:BdryCond} at $\tau=-\epsilon$. This non-hyperbolic metric which should correspond to the solution from path integral optimization according to the higher-dimensional proposal~\eqref{eq:HigherDPathIntAction} has also appeared in the context of the holographic path integral optimization~\cite{Boruch:2021hqs} as a solution to the Neumann boundary condition for a Euclidean planar black hole in $d=4$, dubbed a \emph{Neumann black hole}.\footnote{Interestingly, if one takes the analogy between the two solutions seriously, the constant of integration $\kappa$ should be related to the mass $M$ of the Euclidean planar black hole. The confusing aspect of this analogy is that in the latter case the holographic CFT state should correspond to a TFD state, a fact which did not enter the construction of the $R^{(4)}=$const solution explicitly. }
	
	 In this case, the Q-curvature of the geometry~\eqref{eq:RconstMetric} is given by
	\begin{align}\label{eq:RconstMetricQ4}
		\mathcal{Q}_{4}=6\mu^{2}\left(1-\textrm{sn}^{8}\left( (-1)^{3/4}\,e^{\kappa/2}\,\mu^{1/4}\,\tau+\tau_{0},-1\right)\right)~.
	\end{align}
	The fact that in this case the Q-curvature is not constant can be traced back to the behaviour of the curvature invariant $R_{ab}R^{ab}$, which for a metric of the form~\eqref{eq:CFlat4d} is given by
	\begin{align}\label{eq:RabRab}
		R_{ab}R^{ab}=12e^{-4\phi(\tau)}\left((\partial_{\tau}\phi(\tau))^{4}+(\partial_{\tau}\phi(\tau))^{2}(\partial^{2}_{\tau}\phi(\tau))+(\partial^{2}_{\tau}\phi(\tau))^{2}\right)~,
	\end{align}
	and depends on $\tau$.	
	This shows that in general solving the condition $R=$const in higher-dimensional spacetimes leads to geometries other than the hyperbolic one even after imposing the desired boundary condition at $\tau=-\epsilon$. This result by itself does not contradict the expectations from the higher-dimensional path integral optimization proposal, it does point to the necessity of incorporating more general geometries whose TN interpretation is perhaps less clear.
	
	Furthermore, we can see that imposing $R=$const is a weaker condition than $\mathcal{Q}_{4}=$const. and even in $4$-dimensions, the former can lead to non-hyperbolic metrics consistent with the boundary conditions imposed in the path-integral optimization. On the other hand, the latter condition is seen to lead to hyperbolic metrics for the desired boundary condition and is therefore perhaps more akin to the idea of the emergence of hyperbolic geometries from TN in CFTs.

	\subsection{Tensor Network Interpretation}\label{sec:tn}
	It is interesting to visualize the path integral optimization and generation of the negative curvature from the tensor network renormalization (TNR) perspective. Here, we briefly outline how the notion of penalties arise in the path integral optimization.\footnote{Penalty factors in the context of path integral optimization were previously considered in \cite{Bhattacharyya:2019kvj}.} In the TN approach, one discretizes the path integral over the flat geometry and performs the path integral. The main lesson from the path integral optimization is that one can equivalently perform the path integral over a hyperbolic geometry with a lower cost. In this picture, the cost is associated with the number of tensors in the geometry. The TNR algorithm starts with the coarse-graining in the TN, where one combines $\mathcal{O}(\tau/\epsilon)$ sites over an interval of time $\tau$ and transforms them into a single lattice site~\cite{Caputa:2017yrh}, where $\epsilon$ is the lattice spacing. In other words, the high energy modes $k \gg 1/\tau$, called ``hard sites", are omitted. This is equivalent to penalizing them by a large penalty factor, effectively neglecting them from the geometry. In this way, after the path integral optimization, only the ``easy sites" remain, corresponding to the low energy modes $k \ll 1/\tau$ in the TN.
	
	The following question naturally arises: can we qualitatively estimate the value of the penalty factor for the hard sites in the optimized geometry? To give a rough estimation of this, we follow~\cite{Brown:2016wib, Brown:2017jil}. Consider the optimized $d=2$ metric without setting $\mu = 1$, so that we can consider smooth transition between un-optimized geometry and optimized geometry\footnote{Note that $0 < \mu \leq 1$, where $\mu = 0$ corresponds to the fully un-optimized metric and $\mu = 1$ corresponds to the optimized metric.}
	\begin{align} \label{eq:2dConfMetric2}
		\textrm{d}s^2 = \frac{1}{\mu \tau^2}(\textrm{d} \tau^2 + \textrm{d}x^2).
	\end{align}
	This hyperbolic plane has the curvature length $1/\sqrt{\mu}$ with Gaussian curvature $-\mu < 0$. It is known that the sectional curvature (in $d=2$ the Gaussian curvature coincides with the sectional curvature) needs to be of order $1/K$ (here $K$ is the number of qubits) to generate the negative curvature as well as to see the switchback effect~\cite{Brown:2016wib, Brown:2017jil}. A penalty of order $4^K$ will also lead to the sectional curvature of order $4^K$, thereby requiring a moderate, preferably an $\mathcal{O}(1)$ penalty factor.

This suggests that, in the path integral optimization picture, one can choose the penalty of
	\begin{align}
		p \sim \tilde{\alpha}^{\mathcal{O}(1/|\Lambda|)},  \label{penalty1}
	\end{align}
	to penalize the hard sites in the flat geometry in order to obtain the optimized geometry where $\tilde{\alpha} > 2$ is some number that depends on $K$. Note that this is a version of a progressive penalty considered in~\cite{Auzzi:2020idm}. In other words, in the optimized geometry the hard gates are penalized by a finite amount and there is no need for an infinite penalty to generate the negative curvature. Furthermore, note that the penalty is dependent on the Q-curvature (here the Gaussian curvature) of the optimized geometry. The hard sites in the flat geometry ($|\Lambda| = 0$ i.e., $p \rightarrow \infty$) should be highly penalized, which is consistent with the TN interpretation. The path integral optimization automatically selects the required penalty factor in terms of the Q-curvature. This is highly contrasting with Nielsen's picture, where the penalty factors have to be chosen by hand. This implies, even in the case of complexity geometry \cite{Brown:2017jil, Brown:2019whu}, that the \emph{penalty factor can be chosen according to the underlying geometry}.\footnote{This is analogous to the statement in \cite{Brown:2017jil} that the penalty of $\mathcal{I}_3 >4/3$ make the sectional curvature negative and order $1/K$, we do not need infinite penalty at all. See \cite{Brown:2021euk} for recent works on bounds on complexity choosing penalties.} Moreover, we see that combining Eq.\eqref{eq:RelMuTension} with Eq.\eqref{penalty1} gives an interpretation of the penalty factors from the gravity side. Very naively, the penalty of
	\begin{align}
		p \sim \tilde{\alpha}^{\mathcal{O}(1/(1-T^2))}, \label{penaltygrav}
	\end{align}
	can be interpreted from the gravity side. Thus, we see, on the optimized metric (i.e., $T=0$), the required amount of penalty is finite.
	
	The above analysis strictly holds for $d=2$.  One may wonder whether a similar conclusion can be derived for higher dimensions. In principle one can write a similar un-optimized geometry for higher dimensions where in that case, $\mu$ should be related to the Q-curvature via $\mu = [\Theta_d\,/(d-1)!]^{2/d}$ (see Eq.\eqref{musol0}-Eq.\eqref{musol1}). In such a case, we can intuitively conjecture a penalty factor of the form
		\begin{align}
		p \sim \tilde{\alpha}^{\mathcal{O}\big(\frac{1}{1-T^2/(d-1)^2}\big)},
	\end{align}
	that would penalize the high energy modes of the tensor network. Here we have employed the relation $\mu = 1 - T^2/(d-1)^2$ \cite{Boruch:2020wax}. More quantitative analysis of this proposal would require better understanding of penalty factor in higher-dimensional TN (e.g. extending  \cite{Brown:2021euk} to higher dim.) and we hope to return to this question in future works.
		
\section{Holographic Path Integral Optimization and Higher Curvature on $B$} \label{sec:gra}
In this section we focus on the second question i.e., of higher curvature in the holographic path integral optimization and discuss yet another way that such corrections may enter or be tuned in the path integral optimization. Namely, we perform the optimization be adding by hand (with arbitrary coefficients) higher curvature terms in the induced metric on $B$. At first, including such terms may seem arbitrary and it is not clear at which curvature order one should terminate such procedure. On the other hand, finding $B$ from extremizing an on-shell gravity action with counter-terms computed up to a finite-cutoff region of the bulk is natural in the $T\bar{T}$ context. We discuss this procedure below and point the main difference with the TN ideas based on $T\bar{T}$ deformations \cite{Caputa:2020fbc,Chandra:2021kdv}. 
\subsection{Higher Curvature and Hartle-Hawking Wavefunction}\label{subsec:gracurv}
We first compute a family of Hartle-Hawking wavefunctions discussed before. However, not only with tension $T$ but now with a more general counterterm-like action added on the surface $B$ with arbitrary coefficients. More precisely, we evaluate the classical wavefunction as
	\begin{align} \label{eq:HHAction}
		\Psi_{\textrm{HH}}[\phi]=e^{-I_{\textrm{HH}}[\phi]},\qquad I_{\textrm{HH}}[\phi]=I_G+I_{\textrm{B}},
	\end{align}
	where
	\begin{align} \label{eq:GravityHigherC}
		I_{\textrm{G}}=-\frac{1}{2\kappa^2}\int_{\mathcal{M}} \mathrm{d}^{d+1}x\sqrt{g}\left(R-2\Lambda\right)-\frac{1}{\kappa^2}\int_{\partial\mathcal{M}} \mathrm{d}^{d}x\sqrt{h}K,
	\end{align}
	where $\mathcal{M}$ is the region bounded by $B$ and $\Sigma$, and $\partial\mathcal{M}=B \cup \Sigma$, as in Fig.~\ref{fig:HHSolutionT}. $R$ is the Ricci scalar on region $\mathcal{M}$ and $K$ is the extrinsic curvature on $\partial \mathcal{M}$. Moreover, we take the counterterm-like action on $B$  written in terms of the higher curvature terms as
	\begin{align}\label{eq:BraneAcHigherC}
		I_{\textrm{B}}=\frac{1}{\kappa^2}\int_{B} \mathrm{d}^{d}x\sqrt{h}\left[T+\alpha \mathcal{R}+\beta\mathcal{R}_{ab}\mathcal{R}^{ab} + \gamma\mathcal{R}^2+ \cdots \right],
	\end{align}
	where $\mathcal{R}_{ab}$  and $\mathcal{R}$  denote the Ricci tensor and Ricci scalar of the induced metric on $B$. Note that this is not the exact counterterm action in AdS/CFT, as the coefficients are arbitrary and should be fixed by the optimization\footnote{In principle we should label such multi-parameter HH wave functions by all these coefficients but we avoid this to keep our formulas compact.}.
	
For simplicity, we analyze the vacuum case in Poincar\'{e} AdS$_{d+1}$ coordinates  
	\begin{align}
		\textrm{d}s^2=\frac{\textrm{d}z^2+\textrm{d}x^2_i+\textrm{d}\tau^2}{z^2}~,
	\end{align}
	and consider the region $\mathcal{M}$ contained between the surfaces $z=\epsilon$, denoted as $\Sigma$, and $z=f(\tau)$, denoted as $B$. The induced metric on $B$ is given by
	\begin{align} \label{IndMQP}
		\textrm{d}s^2=\frac{\textrm{d}x^2_i+(1+f'^2)\textrm{d}\tau^2}{f^2}=e^{2\phi(w)}\left(\textrm{d}w^2+\textrm{d}x^2_i\right)~,
	\end{align}
	where we introduced a field $\phi(w)$ and coordinate $w$ as 
	\begin{align}
		e^{2\phi(w)}=\frac{1}{f^2(w)},\qquad w'(\tau)=\sqrt{1+f'(\tau)^2}~.
	\end{align}
	The trace of the extrinsic curvature on $B$ is given by
	\begin{align}
		K_B=-\frac{ff''+d(1-f'^2)}{\sqrt{1-f'^2}}=\frac{e^{-2\phi}\left(\ddot{\phi}+(d-1)\dot{\phi}^2\right)-d}{\sqrt{1-e^{-2\phi}\dot{\phi}^2}}~,
	\end{align}
	in which case the gravity action~\eqref{eq:GravityHigherC} can be directly evaluated yielding \cite{Boruch:2020wax,Boruch:2021hqs}
	\begin{align}
		I_{\textrm{G}}[\phi]&=\frac{V_x(d-1)}{\kappa^2}\int \mathrm{d} w e^{d\phi}\left[\sqrt{1-\dot{\phi}^2e^{-2\phi}} +\dot{\phi}e^{-\phi}\arcsin(\dot{\phi}e^{-\phi})\right]\nonumber \\
		&-\frac{(d-1)}{\kappa^2}\frac{V_xL_\tau}{\epsilon^{d}}-\frac{V_x}{\kappa^2}\left[e^{(d-1)\phi}\arcsin\left(\dot{\phi}e^{-\phi}\right)\right]^0_{-\infty}~.
	\end{align}
	
	Before we go to a more general case, let us first just consider the example where in addition to the tension, we also add the curvature $\mathcal{R}$ with coefficient $\alpha$. With only these two contributions, the action~\eqref{eq:BraneAcHigherC} becomes
	\begin{align}
		I_{\textrm{B}}=\frac{V_x}{\kappa^2}\int \mathrm{d}w\, \left[Te^{d\phi}-\alpha(d-1)e^{(d-2)\phi}\left(2\ddot{\phi}+(d-2)\dot{\phi}^2\right)\right]~,
	\end{align}
	which can be further integrated by parts
	\begin{align}
		I_{\textrm{B}}= \frac{V_x}{\kappa^2}\int \mathrm{d}w\, e^{d\phi}\left[T+\alpha(d-1)(d-2)e^{-2\phi}\dot{\phi}^2\right]- \frac{2\alpha(d-1)V_x}{\kappa^2}\left[e^{(d-2)\phi}\dot{\phi}\right]^0_{-\infty}~.
	\end{align}
Interestingly, this new term not only modifies the bulk equations of motion but also the corner (Hayward) term. The equations of motion arising from the extremisation are given by
	\begin{align}\label{hh1}
		K_B-\frac{d}{d-1}T=-\alpha(d-2)e^{-2\phi}\left(2\ddot{\phi}+(d-2)\dot{\phi}^2\right)~,
	\end{align}
	which can be written as
	\begin{align}
		K_B=\frac{d}{d-1}T+\alpha\frac{d-2}{d-1}\mathcal{R}.
	\end{align}
	This is nothing more than the trace of the general Neumann condition
	\begin{align}
		K_{ij}-K \,h_{ij}=-T \, h_{ij}+2\alpha \mathcal{G}_{ij},
	\end{align}
	where $\mathcal{G}_{ij}=\mathcal{R}_{ij}-\frac{1}{2}\mathcal{R} \,h_{ij}$ is the Einstein tensor written in terms of the brane curvature and the brane metric. If we again look for the solutions of the form
	\begin{align}
		e^{2\phi(w)}=\frac{1}{\mu (w+b)^2}, \label{sol1}
	\end{align}
we obtain the condition between parameters
	\begin{align}
		\frac{T}{d-1}+\sqrt{1-\mu} = \alpha(d-2)\mu. \label{count1}
	\end{align}
It is interesting to note that for $d=2$ the new contribution vanishes and surface $B$ is independent on $\alpha$ i.e., the optimized metric always corresponds to $T=0$.

	Let us then consider $d>2$. Note that $T = -(d-1)$ implies $\mu=0$, corresponding to a fully un-optimized geometry. Suppose now that we want to keep the condition $- (d-1) \leq T \leq 0$. Then,  from Eq.~\eqref{count1} and considering the optimized metric for $\mu=1$ we can solve for the coefficient $\alpha$
	\begin{align}
		\alpha  = \frac{T}{(d-1)(d-2)}~.
	\end{align}
	In order to keep the condition $- (d-1) \leq T \leq 0$, we further require that
	\begin{align}
		- \frac{1}{d-2} \leq \alpha \leq 0.
	\end{align}
	Hence, e.g. the brane action 
	\begin{align}
		I_{\textrm{B}}=\frac{1}{\kappa^2}\int_{B} \mathrm{d}^{d}x\sqrt{h}\left[T + \frac{T}{(d-1)(d-2)} \mathcal{R} \right].
	\end{align}
	will lead to an optimized geometry with $\mu =1$. The lesson from this result is that it is possible to judiciously add a curvature term in the brane action and recover the fully-optimized metric. 
		
	Continuing with this procedure, one can add higher curvature terms on the brane according to~\eqref{eq:BraneAcHigherC}.  After postulating a solution of the form \eqref{sol1}, we then get the following constraint by varying the action with respect to $\phi$ 
	\begin{align}\label{hhh1}
		\frac{T}{d-1}+\sqrt{1-\mu}=(d-2)\alpha \mu-(d-4)(d-1)\mu^2(\beta+d\gamma).
	\end{align}
Again, the higher-dimensional contributions identically vanish in $d=4$. If we are interested in $d>4$, then we can solve this constraint by e.g. taking $\alpha,\beta,\gamma$ as
	\begin{align}\label{eq:count4}
		\alpha=\frac{T}{(d-1)(d-2)},\qquad \beta = -d\gamma,\qquad \gamma=T,
	\end{align}
however, there are many other choices that will equivalently lead to the optimized geometry with $\mu=1$. At the moment we do not have a strong argument to resolve this ambiguity and we hope that better understanding of the role of Q-curvature action in the CFT optimization may help in this task.

More generally, we can think about the above procedure as follows. When we vary the on-shell action with general series of counter-terms with respect to the induced metric on $B$, we compute
\begin{equation}
\delta I_{HH}\sim T_{ij}h^{ij},
\end{equation}
where $T_{ij}$ (denoted in analogy with the holographic stress-tensor \cite{Balasubramanian:1999re}) is generally
\begin{equation}
T_{ij}\sim K_{ij}-Kh_{ij}+t_{ij},
\end{equation}
and $t_{ij}$ comes from the variation of the additional terms on $B$. If we set this variation to $0$, we impose the Neumann boundary condition on $B$, \emph{i.e.}, $T_{ij}=0$. Moreover, in our gauge $h_{ij}=e^{2\phi}\delta_{ij}$, the condition that variation of $I_{HH}$ with respect to $\phi$ vanishes corresponds to
\begin{equation}
T^i_i=0\quad  \Leftrightarrow\quad K=\frac{1}{d-1}t^i_i.
\end{equation}
If we only work in pure gravity (as in \cite{Boruch:2020wax,Boruch:2021hqs}) and add ``geometric" terms on $B$, the Hamiltonian constraint gives the condition
\begin{equation}
\mathcal{R}=2\Lambda+K^2-K_{ij}K^{ij}=2\Lambda+\frac{1}{d-1}(t^i_i)^2-t_{ij}t^{ij}.
\end{equation}
For example, for only the tension term on $B$, we have $t_{ij}=T h_{ij}$ and the Ricci scalar on $B$ is constant negative. Similarly, with higher curvature terms, we can find our solution \eqref{sol1} that also has negative curvature $\mathcal{R}=-d(d-1)\mu$ where $\mu$ is expressed in terms of $T$, $\alpha$ and the other parameters via the maximization equation.
\subsection{$T\bar{T}$ and Holographic Path Integral Optimization }\label{TT-bar}
Let us now discuss the connection to the so-called $T\bar{T}$-deformations. According to the proposal of \cite{McGough:2016lol}, we could interpret the bulk on-shell action with Dirichlet boundary condition on $B$ as an effective holographic description of a CFT deformed by the higher-dimensional $T^2$ operator \cite{Hartman:2018tkw}. Given the interpretation that the holographic path integral optimization should be thought of as the boundary action plus finite cut-off terms, it  is tempting to speculate that, in holographic settings, $T\bar{T}$ deformations could be used as a tool to introduce such finite cut-off corrections (see \cite{Caputa:2020fbc,Chandra:2021kdv}). Making this precise is beyond the scope of this work however we discuss below how these two approaches may be mutually consistent.

More precisely, the effective gravity action that describes a $T^2$-deformed holographic CFT is given by
\begin{equation}
I_{T\bar{T}}=-\frac{1}{2\kappa^2}\int\sqrt{g}\left(R-2\Lambda\right)-\frac{1}{\kappa^2}\int \sqrt{h}K+S_{ct}
\end{equation}
where the appropriate holographic counter-terms integrated up to the cut-off surface $B$ are\footnote{We use the standard holographic counter-term up to $d=6$.}
\begin{equation}
S_{ct}=\frac{1}{\kappa^2}\int\sqrt{h}\left[(d-1)+\frac{\mathcal{R}}{2(d-2)}+\frac{(\mathcal{R}_{ij}\mathcal{R}^{ij}-\frac{d}{4(d-1)}\mathcal{R}^2)}{2(d-4)(d-2)^2}\right].
\end{equation}
Once we compute the holographic stress tensor from this action, solve it for $K_{ij}$ and $K$ (in terms of $T_{ij}$ and $T^i_i$), the Hamiltonian constraint of gravity can be written as the anomaly equation 
\begin{equation}
\langle \hat{T}^{i}_{\phantom{i}i}\rangle=-\frac{1}{16\pi G_N}\mathcal{R}-4\pi G_N\left(\hat{T}_{ij}\hat{T}^{ij}-\frac{1}{d-1}(\hat{T}^i_i)^2\right),
\end{equation}
where $\hat{T}_{ij}=T_{ij}+a_d t_{ij}$ is the appropriately renormalised (by the counter-terms) holographic stress tensor. In $d=2$ this relation is simply the anomaly equation together with the $T\bar{T}$ operator but in higher (even) dimensions one can also separate holographic anomalies (e.g.  in $d=4$ with central charges $a=c$) and the remaining part define the $T^2$ operators on curved background in the holographic large-N regime \cite{Hartman:2018tkw,Caputa:2019pam}. 

If we would naively minimize this action with respect to the choice of the induced metric on $B$ this would be equivalent to setting $T_{ij}$ to zero. But this is not what is being done in the $T\bar{T}$ (or $T^2$ in higher-dimensional) TN~\cite{Caputa:2020fbc}. There, we would simply consider constant mean curvature slices $B$ with a non-trivial stress tensor i.e., Dirichlet boundary condition on $B$. On the other hand, in the path integral optimization, we fix $B$ by imposing the Neumann boundary condition on these slices. Still the two approaches can be consistent and give rise to the same slices of the bulk (see e.g. \cite{Caputa:2020fbc}) that have a constant Ricci scalar $\mathcal{R}$. A precise understanding of the relation between these two constructions may involve some version of the Legendre transform that has been discussed in the context of the $T\bar{T}$ deformation in \cite{Coleman:2020jte} (based on \cite{Witten:2018lgb}) and we leave this as an exciting future problem.

\subsection*{Acknowledgements}
We thank Jan Boruch and Dongsheng Ge for discussions and Tadashi Takayanagi for important comments on the draft. We also thank Aninda Sinha for participation in the initial stages of this project and helpful suggestions. PN thanks the organizers of ``Quantum Information in QFT and AdS/CFT-II" where a part of the work was presented. HC is partially supported by the International Max Planck Research School for Mathematical and Physical Aspects of Gravitation, Cosmology and Quantum Field Theory and by the Gravity, Quantum Fields and Information (GQFI) group at the Max Planck Institute for Gravitational Physics (Albert Einstein Institute). The GQFI group is supported by the Alexander von Humboldt Foundation and the Federal Ministry for Education and Research through the Sofja Kovalevskaja Award. PN acknowledges the University Grants Commission (UGC), Government of India, for providing financial support. The work of P.C. is supported by NAWA “Polish Returns 2019” and NCN Sonata Bis 9 grants.

	\bibliographystyle{JHEP}
	\bibliography{references}     
	
\end{document}